\documentclass{article}
\usepackage{microtype}
\usepackage{tabularx}
\usepackage[section]{placeins}
\usepackage{graphicx,subfigure}
\usepackage{authblk}
\usepackage{amsmath}
\usepackage{amssymb}
\usepackage[dvipsnames]{xcolor}
\usepackage{tikz}
\usepackage{float}
\usepackage{dsfont}
\usepackage[title]{appendix}
\usepackage[margin=1in]{geometry}
\usepackage{xurl}
\usepackage{hyperref}
\usetikzlibrary{positioning}
\usetikzlibrary{shapes.geometric}
\usepackage[backend=biber,style=vancouver]{biblatex}
\begin{document}
\title{\Large A model of tuberculosis progression using CompuCell3D}
\author[1,*]{James W. G. Doran}
\author[2]{Christopher F. Rowlatt}
\author[3]{Gibin G. Powathil}
\author[1]{Ruth Bowness}
\author[1]{Christian A. Yates}
\affil[1]{Department of Mathematical Sciences, University of Bath, Bath, United Kingdom BA2 7AY}
\affil[2]{Institute for Mathematical Innovation, University of Bath, Bath, United Kingdom BA2 7AY}
\affil[3]{Department of Mathematics, Faculty of Science and Engineering, Swansea University, Swansea, United Kingdom SA1 8EN}
\affil[*]{Corresponding author: James W. G. Doran, jd521@bath.ac.uk}
\maketitle
\begin{abstract}
    Tuberculosis (TB) is an airborne disease caused by the bacterium \textit{Mycobacterium tuberculosis} (\textit{M. tb}). Prior to the COVID-19 pandemic, TB was the leading cause of death from an infectious agent globally. However, most people exposed to \textit{M. tb} do not develop active TB and go on to display symptoms. Instead, in the majority of cases, the bacteria are contained within a granuloma (an aggregation of immune cells) without being eliminated; this is called latent TB. The spatial organisation of the bacteria and immune cells is important in determining whether an individual exposed to \textit{M. tb} will develop latent or active TB.\par
    In this paper, we present a multi-cell, multiscale model of TB progression to investigate the importance of the spatial organisation. This is a novel TB within-host dynamics modelling framework, having been developed using \texttt{CompuCell3D} (CC3D), an open-source computer software used for simulating cellular biological processes both within and between cells. We used this model to compare the generated results with those from a previously developed within-host infectious disease model. We found that, although the results of our CC3D model mostly agree qualitatively with those from the previously developed model, there are quantitative differences. Additionally, we conducted a robustness analysis of key model parameters from the CC3D model to determine their importance to the CC3D model output, using a methodology specifically designed for agent-based models. The model output appears to be robust in response to perturbations in parameters controlling chemotactic movement, but less so in response to perturbations in parameters controlling persistence of movement in cells, cell adhesion and volume constraints. This work compares our CC3D model of TB progression with another agent-based modelling approach to the same problem.
\end{abstract}
\section*{Keywords}
\texttt{CompuCell3D}, tuberculosis, bacteria, cellular Potts model, within-host
\section{Introduction}
\label{section: Introduction}
%FIRST PARAGRAPH - CONTEXT
Tuberculosis (TB) is an airborne disease caused by \textit{Mycobacterium tuberculosis} (\textit{M. tb}) \cite{WHO2022}. It is transmitted to susceptible humans through inhalation after being expelled into the air from an infectious host via aerosols (see \cite{Tellier2019} and references therein). This expulsion of aerosols containing \textit{M. tb} can potentially occur after any respiratory activity \cite{Patterson2019}. Although TB can affect multiple organs, the site of infection is most commonly located within the lungs \cite{WHO2022}. According to the 2025 Global Tuberculosis Report published by the World Health Organisation (WHO), TB is the leading cause of death from an infectious agent and is one of the top 10 causes of death worldwide \cite{WHO2025}. Despite this, most individuals (approximately $90\pm 5\%$) exposed to \textit{M. tb} do not become ill \cite{Ahmad2011,Glaziou2018}. It is unknown if some of these individuals are able to eradicate the bacteria entirely \cite{Glaziou2018}; it is estimated that most of them (again, approximately 90\%) will experience a latent version of TB \cite{Ahmad2011}. Those who do become ill and suffer from active TB are usually treated with a combination of antibiotics; the standard regimen for drug-susceptible TB lasts six months and includes four antibiotics \cite{WHO2022}. A failure to complete a full regimen can lead to antibiotic-resistant strains emerging \cite{WHO2022}, with multidrug-resistant/extensively drug-resistant TB (MDR/XDR-TB) patients needing longer and more complex treatment protocols \cite{Falzon2011}. Although attempts have been made to reduce treatment duration (e.g., \cite{Gillespie2014}), these have so far proven unsuccessful.\par
%SECOND PARAGRAPH - NEED
Efforts have been made to understand the within-host disease dynamics of TB, and why people have different responses to the \textit{M. tb} pathogen (including the models in \cite{Wigginton2001}, \cite{Gammack2004}, \cite{Marino2004}, \cite{SegoviaJuarez2004}, \cite{Ganguli2005} and \cite{Pitcher2020}). To help determine the factors that lead to TB relapses, \citeauthor{Bowness2018} (\citeyear{Bowness2018}) developed an agent-based model (ABM) of the within-host progression of TB \cite{Bowness2018}. One of the main conclusions of the paper was that the spatial location of the initial bacterial cluster and the arrangement of the blood vessel sources within the lung tissue was important, as the immune response required to contain the outbreak (and the drugs needed if an individual did go on to develop active TB) could reach the site of infection more quickly if this cluster was close to the blood vessels from which the immune cells (and drugs) originated. This suggests that the spatial organisation of \textit{M. tb} bacteria and immune cells is an important factor in determining whether an individual exposed to the pathogen will experience latent or active TB.\par
%THIRD PARAGRAPH - TASK
We developed a similar model of the first 200 hours of within-host TB progression, using \texttt{CompuCell3D} \cite{Swat2012}, in order to compare findings and modelling approaches against the model of \citeauthor{Bowness2018} \cite{Bowness2018}. \texttt{CompuCell3D} software is used for modelling biological cellular processes, primarily using a Cellular Potts approach (for more information, see \cite{Glazier2007}) to represent cell motility on a lattice, subject to adhesion rates that differ between different types of neighbouring cells and volume constraints. The agent-based model components are combined with partial differential equation (PDE) solvers capturing the reaction-diffusion processes of any interacting chemical fields. Subcellular processes can be incorporated using ordinary differential equations (ODEs). Other cell behaviours, such as chemotaxis (directed movement towards, or away from, high chemical concentrations) and secretion and absorption of different chemicals, can be included, with the possibility of changing cell properties partway through simulations.\par
Prior to the release of \texttt{CompuCell3D}, the Glazier-Graner-Hogeweg model (equivalent to the Cellular Potts model) was previously used to investigate cell sorting \cite{Graner1992}, the migration and thermotaxis of \textit{Dictyostelium discoideum} slugs \cite{Maree1999},
and the evolutionary dynamics that lead to morphogenesis \cite{Hogeweg2000}, among other areas of study. Since its release, \texttt{CompuCell3D} has been used to model a range of biological cellular phenomena, such as angiogenesis, tumour growth (both modelled in \cite{Shirinifard2009}), and subcellular behaviours \cite{Hester2011}. In particular, a number of infectious diseases have been studied using this modelling framework, including COVID-19 \cite{FerrariGianlupi2022} and influenza \cite{Sego2022}. However, to the best of our knowledge, ours is the first within-host model of TB progression developed using \texttt{CompuCell3D}.\par
%FOURTH PARAGRAPH - OBJECT
This work will be insightful for any researchers wishing to develop mathematical models of cellular-level processes, specifically of within-host infectious disease dynamics, using this software in future, as it will provide a useful framework for comparing \texttt{CompuCell3D} with other ABMs of within-host infectious disease dynamics.\par
The rest of this paper is structured as follows. Section \ref{sec: host-pathogen} describes the host-pathogen interactions, providing a broad overview of the main biological components and processes; Section \ref{section:Methods} describes the specific details of the model developed in \texttt{CompuCell3D}, and the conceptual ideas behind it; Section \ref{section:Results} compares the output of 300 simulations of the model against the output of 300 simulations from the Within-Host Infectious Disease Model (hereafter referred to as WHIDM \cite{Bowness2018}), an alternative agent-based modelling framework adapted from the modelling framework of \citeauthor{Bowness2018} \cite{Bowness2018}\footnote{It should be noted that, in contrast to the simulations run by \citeauthor{Bowness2018} in \cite{Bowness2018}, we only simulate the first 200 hours post-infection on WHIDM, and include neither antibiotics nor an adaptive immune response. This is to ensure the comparison is as close as possible between WHIDM and the CC3D model presented in this article.}. The determination of the number of simulations required is outlined in Section \ref{section:Consistency analysis}. Section \ref{section:Results} also details the findings of a robustness analysis of key model parameters; Section \ref{section:Discussion} summarises the findings of this paper, highlights the differences between the two modelling frameworks considered, and outlines planned future work.
\subsection{Host-pathogen interactions}
\label{sec: host-pathogen}
The interactions described in this section are the ones considered by WHIDM \cite{Bowness2018}; in reality, the biology is more complicated than what is presented here, and we have excluded some biological mechanisms from the CC3D model to replicate WHIDM as closely as possible. For example, a range of cytokines are known to impact within-host TB dynamics, such as IFN-$\gamma$, IL-4, IL-10 and IL-12 \cite{Wigginton2001}, and these perform a number of different functions, such as up-regulation of certain immune response and down-regulation of others. However, both the CC3D and WHIDM models presented here use a single cytokine, which is a generic chemokine used for chemotaxis of immune cells.\par
\textit{M. tb} cells depend on the local oxygen concentration to replicate, but are capable of switching to a `dormant' or slower-growing state if there is an insufficient oxygen supply \cite{Lipworth2016} \cite{Hammond2015}. Immune cells, such as macrophages and T-cells, rely upon chemokines to direct them to the site of infection. They follow the chemical signals of the chemokines via chemotaxis: directed movement according to a chemoattractant gradient (in this case, towards an increasing chemoattractant concentration).\par
The immune system has two potential stages of its response to this infection: the innate immune response, and the adaptive immune response. In our model, the innate immune response to the bacteria is carried out by macrophages. Upon discovering the bacteria, they ingest them in a process known as phagocytosis. Additionally, they secrete chemokines to recruit more macrophages to the site of infection. If enough macrophages encounter and remove the bacteria from the domain, the infection is contained. However, each infected macrophage has a finite carrying capacity for intracellular bacteria; once this carrying capacity is breached, they burst and release their intracellular bacteria back into the tissue. They leave behind necrotic material, known as caseum.\par
If the innate immune response fails to contain the outbreak, the adaptive immune response is the second stage of the immune system's attempt to control the infection. A signal, in the form of diffusing chemokines released by infected macrophages, is sent to the lymph nodes recruiting T-cells to the site of infection. Upon receiving this signal, T-cells arrive at the infected lung tissue (around 9 to 11 days after the initial infection \cite{Urdahl2011}) and initiate the adaptive immune response. There are different types of T-cells, each carrying out different functions, including killing infected macrophages by triggering apoptosis (a form of programmed cell death). We have chosen not to include T-cells in our model, as their response does not typically begin until after the first 200 hours post-infection that we are investigating \cite{Urdahl2011}. Thus, we did not add T-cells to the CC3D model, and we removed them from the simulations run using WHIDM. A pathologic hallmark of TB is the formation of granuloma structures \cite{Ramakrishnan2012}, where an aggregation of bacteria, immune cells and caseum form at the site of an infection. The diffusion of chemical species, such as oxygen and cytokines, through a granuloma are known to be reduced in comparison to normal lung tissue \cite{Datta2016} \cite{Pienaar2016}.\par
The immune response can then either completely eradicate the bacteria, containing the bacteria within one or more granulomas (leading to latent TB) or fail to contain the infection leading to dissemination of the infection (active TB). In the case of containment, granulomas will typically be small and well-defined; any bacteria left will be trapped within their ``walls". On the other hand, if granulomas are large and not well-defined, bacteria can escape, and this will often lead to dissemination. Having said this, it can be difficult to predict whether or not an individual will progress to active TB \cite{Phillips2016}. Additionally, liquefaction of solid caseum into a liquid state has been implicated in the progression to active TB \cite{Cardona2011} and relapse of disease \cite{Bowness2018}. Our modelling framework explores the evolution of a single TB granuloma, and whether it is likely to result in containment or dissemination; the details of this are presented in the following section.
\section{\texttt{CompuCell3D} computational model}
\label{section:Methods}
\begin{figure}
    \centering
    \resizebox{10cm}{!}{
    \begin{tikzpicture}
        \node[draw] (vessels) at (-5,2){Vessel distribution};
        \node[draw] (oxygen) at (-5,0){Oxygen supply};
        \node[draw,text width = 4.5cm] (oxygenPDE) at (-5,-2){Diffusion of oxygen (PDE) $\rightarrow$ spatial distribution of oxygen};
        \node[draw, text width = 2.5 cm] (immuneCells) at (0,0){Location and number of immune cells (ABM)};
        \node[draw,text width = 4.5cm] (cytokinePDE) at (5,2){Diffusion of cytokines (PDE) $\rightarrow$ spatial distribution of cytokines};
        \node[draw] (cytokines) at (5,0){Cytokine supply};
        \node[draw,text width = 3cm] (bacteria) at (0,-2){Fast- and slow-growing bacteria (ABM)};
        \draw[-stealth] (vessels.south) -- (oxygen.north);
        \draw[-stealth] (oxygen.south) -- (oxygenPDE.north);
        \draw[-stealth] (oxygenPDE.east) -- (bacteria.west);
        \draw[-stealth] (immuneCells.south) -- (bacteria.north);
        \draw[-stealth] (immuneCells.east) -- (cytokines.west);
        \draw[-stealth] (cytokines.north) -- (cytokinePDE.south);
        \draw[-stealth] (cytokinePDE.west) -| (immuneCells.north);
    \end{tikzpicture}
    }    
    \caption{Schematic describing the basic processes of the model (modified from \cite{Bowness2018}). An arrow from box A to box B indicates box A impacts upon box B. Abbreviations: PDE, partial differential equation; ABM, agent-based model.}
    \label{fig:CC3D schematic}
\end{figure}
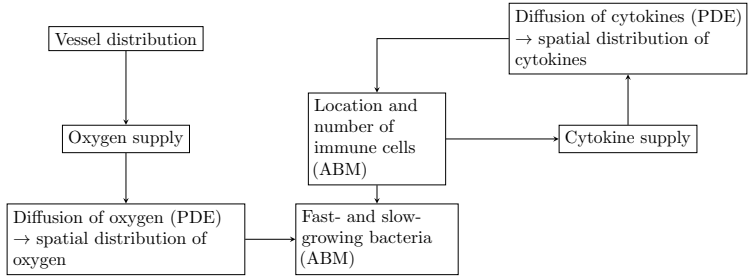

This section outlines the computational model (developed using \texttt{CompuCell3D}) which we use to simulate TB progression in lung tissue. It should be noted that the CC3D model has been developed to mimic the mechanisms used in the WHIDM framework as closely as possible, in order to compare the two models. The rest of this section is structured as follows: Section \ref{sec:model_env} describes the model environment; Section \ref{sec:sim} discusses the general properties of each repeat of the model; Section \ref{sec:components} provides details about the different biological cells and fields represented in our model; Section \ref{sec:processes} lays out the processes that occur throughout a simulation; Section \ref{sec:params} explains the rationale behind the chosen parameter values; and Section \ref{sec: comparison to WHIDM} compares the methodology of the CC3D model and WHIDM. The basic processes of the conceptual model are summarised in the schematic in Figure \ref{fig:CC3D schematic}.
\subsection{Model environment}
\label{sec:model_env}
The model represents a $2\text{mm} \times 2\text{mm}$ two-dimensional cross-section of lung tissue. $N_{bv}$ blood vessels and $M^r_{init}$ macrophages are initially randomly located throughout the tissue; the blood vessels remain fixed in position whilst the macrophages move through the tissue over time. It has previously been suggested that a minimal infectious dose of \textit{M. tb} is of the order of 10 to 100 bacteria \cite{CapuanoIII2003}. On this basis, we initialise the model with an initial cluster of 12 \textit{M. tb}, 6 fast-growing and 6 slow-growing, in-keeping with \cite{Bowness2018}. To do this, the lower bounds for a square region of a fixed size are generated uniformly at random from the $x$- and $y$- axes, and the upper bounds are set to be a fixed amount larger than the lower bounds. As such, the smallest possible lower bound is 0 and the largest possible lower bound causes the upper bound to equal the maximum value on the axis in question. Within this square region, the $x$- and $y$-coordinates for each bacterium are chosen uniformly at random, such that at most one bacterium occupies each lattice site. These bacteria replicate over time, with the rate of replication dependent upon the amount of oxygen, supplied by blood vessels, available to the bacteria. More detail on how \textit{M. tb} replication is handled by our model is given in Section \ref{sec:replication}.\par
\subsection{General simulation properties}
\label{sec:sim}
The simulation accounts for time in terms of Monte Carlo Steps (MCS), where a Monte Carlo Step is $N^2$ lattice copy attempts. Here, $N^2$ is equal to the number of voxels in the domain of size $N \times N \times 1$: as we wish to model in two dimensions, we have set the length of the third dimension, required by the CC3D framework, equal to 1.\footnote{As our simulation is in 2D, we will refer to voxels as pixels going forward.} Each MCS represents 24s of real time: for justification, see the paragraph on parameter estimation at the end of Section \ref{sec: GGH}. We let $T \in \mathbb{R}$ denote the length of the simulation in Monte Carlo steps (each Monte Carlo Step $(t,t+\Delta t]$ lasts $\Delta t$, equivalent to 24s of real time, where the time at the start of the Monte Carlo Step $t \in \{0,\Delta t,...,T-\Delta t\}$) and $I = [0,T) \subset \mathbb{R}$ denote the time interval which comprises the total run time of the simulation. Furthermore, we will denote each lattice site as $\vec{i} = (l,m), l,m \in \{1,...,N\}$, the index of each lattice site $\vec{i}$ as $\sigma(\vec{i})$, and the type of cell at lattice site $\vec{i}$ as $\tau(\sigma(\vec{i}))$. Each cell is a set of adjacent lattice sites with a shared index $\sigma$ \cite{Swat2012}.
\subsection{Biological components}
The biological components represented in this model are several cell types, and a variety of chemical fields. In Section \ref{sec:cells}, we outline the cell types represented in our model, and in Section \ref{sec:fields}, we discuss the chemical fields. Model variables will be given in typewriter font in what follows.
\label{sec:components}
\subsubsection{Cells}
\label{sec:cells}
Each CC3D model has objects that live on the lattice sites; these fundamental objects are known as ``generalised cells'' \cite{Swat2012}. Each generalised cell can be a biological cell, a subcellular compartment, a cluster of cells, non-cellular material or surrounding medium \cite{Swat2012}. Our model contains four main classes of generalised cells: blood vessels (\texttt{BV}), \textit{Mycobacterium tuberculosis} (\textit{M. tb}), macrophages, and necrotic material, referred to here as caseum (\texttt{CASEUM}). The \textit{M. tb} can be in one of three states: fast-growing extracellular (\texttt{FGEB}), slow-growing extracellular (\texttt{SGEB}), and intracellular. We do not explicitly represent the intracellular bacteria in our model, in keeping with other agent-based models of within-host TB dynamics, e.g. \cite{Bowness2018}; instead, any infected/chronically infected macrophage has a counter which keeps track of the number of bacteria they have phagocytosed. The macrophages can also be in one of three states: resting (\texttt{RM}), infected (\texttt{IM}), and chronically infected (\texttt{CIM}). Additionally, a \texttt{WALL} cell type is included in our CC3D model. \texttt{WALL} cells are used to define a thin layer of cells surrounding the boundary of the domain. They prevent other CC3D cells from sticking to the boundary and enforce no-flux boundary conditions; essentially, the positions of the \texttt{WALL} cells are equivalent to the boundary of the domain \cite{Swat2018} (see \cite{Swat2009} for an example of a CC3D model developed with a \texttt{WALL} cell type to enforce no-flux boundary conditions).\par
Symbolically, we can label the sets of different cell types as follows:
\begin{itemize}
    \item $\Omega^{bv} = \bigcup_{j=1}^{N_{bv}} \Omega_j^{bv}$ represents the set of blood vessels, where $\Omega_j^{bv}$ is the $j^{\text{th}}$ blood vessel, $j=1,...,N_{bv}$ (that is, each $\Omega_j^{bv}$ represents the set of lattice sites $\vec{i}$ that make up a blood vessel).
    \item $\Omega^{rm} = \bigcup_{j=1}^{N_{rm}} \Omega_j^{rm}$ represents the set of resting macrophages, where $\Omega_j^{rm}$ is the $j^{\text{th}}$ resting macrophage, $j=1,...,N_{rm}$.
    \item $\Omega^{im} = \bigcup_{j=1}^{N_{im}} \Omega_j^{im}$ represents the set of infected macrophages, where $\Omega_j^{im}$ is the $j^{\text{th}}$ infected macrophage, $j=1,...,N_{im}$.
    \item $\Omega^{cim} = \bigcup_{j=1}^{N_{cim}} \Omega_j^{cim}$ represents the set of chronically infected macrophages, where $\Omega_j^{cim}$ is the $j^{\text{th}}$ chronically infected macrophage, $j=1,...,N_{cim}$.
    \item $\Omega^f = \bigcup_{j=1}^{N_f} \Omega_j^f$ represents the set of fast-growing extracellular bacteria, where $\Omega_j^f$ is the $j^{\text{th}}$ fast-growing extracellular bacterium, $j=1,...,N_f$.
    \item $\Omega^s = \bigcup_{j=1}^{N_s} \Omega_j^s$ represents the set of slow-growing extracellular bacteria, where $\Omega_j^s$ is the $j^{\text{th}}$ slow-growing extracellular bacterium, $j=1,...,N_s$.
    \item $\Omega^c = \bigcup_{j=1}^{N_c} \Omega_j^c$ represents the set of caseous cells, where $\Omega_j^c$ is the $j^{\text{th}}$ caseous cell, $j=1,...,N_c$.
\end{itemize}
Each of these generalised cells can comprise multiple lattice sites, but no lattice site can be occupied by more than one cell type. Any lattice site that is not occupied by one of these cell types is instead occupied by the surrounding medium, which is represented by a special generalised cell type, \texttt{Medium} - this cell type is required in all \texttt{CompuCell3D} models. We will define $\Omega^M$ to be the set of lattice sites occupied by \texttt{Medium}. To simplify, we define $\Omega^m = \Omega^{rm} \cup \Omega^{im} \cup \Omega^{cim}$ to be the set of macrophages, and $\Omega^b = \Omega^f \cup \Omega^s$ to be the set of extracellular bacteria. We can then say that $\Omega = \Omega^{bv} \cup \Omega^b \cup \Omega^m \cup \Omega^c \cup \Omega^M \cup \partial \Omega$, where $\partial \Omega$ is the boundary of the domain and is comprised of the \texttt{WALL} cells. In other words, $\Omega$ is the set of all lattice sites $\vec{i}=(l,m),l,m \in \{1,...,N\}$. All the cell types used in our model and their respective indices are given in Table \ref{tab:cell types and indices}: these are the values for $\tau(\sigma(\vec{i}))$ when cell types appear in later equations in this article.\par
\begin{table}
    \centering
    \begin{tabular}{|l|l|l|}
        \hline
        \textbf{Cell type} & \textbf{Index} & \textbf{Frozen in place?}\\
        \hline
        Medium & 0 & \\
        Blood vessel (BV) & 1 & \checkmark\\
        Fast-growing extracellular bacterium (FGEB) & 2 & \\
        Slow-growing extracellular bacterium (SGEB) & 3 & \\
        Resting macrophage (RM) & 4 & \\
        Infected macrophage (IM) & 5 & \\
        Chronically infected macrophage (CIM) & 6 & \\
        Caseum & 7 & \checkmark\\
        Wall & 8 & \checkmark\\
        \hline
    \end{tabular}
    \caption{All cell types and their respective numerical type indicators included in the simulation.}
    \label{tab:cell types and indices}
\end{table}
The intracellular bacteria do not have a CC3D cell type; any intracellular bacteria are instead tracked by counters within the infected macrophages and chronically infected macrophages that phagocytosed them. When an extracellular bacterium is phagocytosed, it is removed from the model; if the macrophage that phagocytosed it bursts or dies naturally, which would release the bacterium (and any others that were phagocytosed by that macrophage), the bacterium is allowed to grow extracellularly once more, in a slow-growing state (having been starved of oxygen within the macrophage), so new \texttt{SGEB} cells are created, one for each released bacterium, to represent this. At the point of the macrophage bursting or dying naturally, neighbouring lattice sites are checked to see if they are unoccupied (that is, occupied by the \texttt{Medium} cell type). As an attempt to reduce lattice anisotropy (that is, the tendency of cells to be biased towards moving parallel to the directions of the lattice axes) \cite{Swat2012}, we increased the neighbour order to 5, hence we sum over the nearest neighbours, second-nearest neighbours and so on up to fifth-nearest neighbours of each of the lattice sites $\vec{i}$ in the macrophage that is bursting or dying naturally (see \cite{Swat2009} for more details on neighbour order). If a lattice site is found to be unoccupied, one of the phagocytosed bacteria being released is generated at this lattice site. If no neighbouring lattice sites are unoccupied for such a  bacterium to be generated, the bacterium dies and is not generated after the macrophage bursts or dies naturally, neither in the current time step nor in future time steps.
\subsubsection{Chemical fields}
\label{sec:fields}
In our model, two chemical fields are included: \texttt{Oxygen} (represented hereafter as $O$); and a chemokine molecule to attract macrophages to the site of infection, \texttt{Chemokine} (represented hereafter as $C$). These are both represented as solutions of partial differential equations (PDEs), which evolve over space and time and incorporate terms to cover secretion, absorption, diffusion and decay, where applicable. The PDEs which give rise to these solutions are outlined further in Sections \ref{sec:oxygen} and \ref{sec:cytokines}.
\subsection{Biological processes}
\label{sec:processes}
\subsubsection{The Glazier-Graner-Hogeweg model}
\label{sec: GGH}
\texttt{CompuCell3D} is software which utilises the Glazier-Graner-Hogeweg model (also known as the Cellular Potts model: for more information on this model, see \cite{Glazier2007}). The model evolves over a series of MCS (defined in Section \ref{sec:sim}). Each lattice site is assigned an index according to the type of cell which occupies it at that particular time. In a given MCS, for each lattice site $j \in \{1,...,N^2\}$, a neighbouring lattice site $i$ is randomly chosen, and an attempt is made to copy the index of the cell type at site $j$ to site $i$. This will lead to a new proposed configuration of cells. To decide whether or not this new configuration of cells is accepted or not, the total energy of the current configuration is computed and compared to the energy of the new proposed configuration, were it to be accepted.  The total energy of a certain configuration of cells is calculated as the energy required for cells to adhere to one another, plus the energy associated with cells attempting to remain at their preferred volume, plus the kinetic energy required for both targeted movement (chemotaxis is the only type of targeted movement exhibited by the cells in our model) and random movement.\par
The equation for calculating the total energy of a configuration of cells is as follows:
\begin{equation}
\label{eq:hamiltonian}
    H = E_{\text{adhesion}} + E_{\text{volume}} + E_{\text{chem}} + E_{\text{vec}}.
\end{equation}
In Equation (\ref{eq:hamiltonian}), $H$ is the total energy of the configuration of cells, known as the Hamiltonian, $E_{\text{adhesion}}$ is the total energy required for the current cell adhesions (see Equation (\ref{eq:adhesion})), $E_{\text{volume}}$ is the total energy used by the cells to satisfy their volume constraints (see Equation (\ref{eq:volume})), $E_{\text{chem}}$ is the total kinetic energy of cells moving due to chemotaxis (see Equation (\ref{eqn:chemotaxis})), and $E_{\text{vec}}$ is the energy associated with random movement: in the absence of chemotaxis, this will dictate the diffusion of cells (see Equation (\ref{eq:direction})).\par
\begin{figure}
    \centering
    \resizebox{10cm}{!}{
        \begin{tikzpicture}
            \filldraw[fill=Black] (-5,-2) rectangle (-1,2);
            \filldraw[fill=Black] (1,1) rectangle (5,5);
            \filldraw[fill=Black] (1,-5) rectangle (5,-1);
            \draw[style=help lines,draw=White] (-5,-2) grid[step=0.5cm] (-1,2);
            \draw[style=help lines,draw=White] (1,1) grid[step=0.5cm] (5,5);
            \draw[style=help lines,draw=White] (1,-5) grid[step=0.5cm] (5,-1);
            \filldraw[fill=Red,draw=Red] (-3.5,1) rectangle (-2.5,1.5);
            \filldraw[fill=Red,draw=Red] (-4,-1) rectangle (-2,1);
            \filldraw[fill=Red,draw=Red] (-3.5,-1.5) rectangle (-2.5,-1);
            \filldraw[fill=Red,draw=Red] (-4.5,-0.5) rectangle (-4,0.5);
            \filldraw[fill=Red,draw=Red] (-2,-0.5) rectangle (-1.5,0.5);
            \filldraw[fill=Red,draw=Red] (2.5,4) rectangle (4,4.5);
            \filldraw[fill=Red,draw=Red] (2,2) rectangle (4,4);
            \filldraw[fill=Red,draw=Red] (2.5,1.5) rectangle (3.5,2);
            \filldraw[fill=Red,draw=Red] (1.5,2.5) rectangle (2,3.5);
            \filldraw[fill=Red,draw=Red] (4,2.5) rectangle (4.5,3.5);
            \filldraw[fill=Red,draw=Red] (2.5,-2) rectangle (3.5,-1.5);
            \filldraw[fill=Red,draw=Red] (2,-4) rectangle (4,-2);
            \filldraw[fill=Red,draw=Red] (2.5,-4.5) rectangle (3.5,-4);
            \filldraw[fill=Red,draw=Red] (1.5,-3.5) rectangle (2,-2.5);
            \filldraw[fill=Red,draw=Red] (4,-3.5) rectangle (4.5,-2.5);
            \draw[-stealth] (-3,2) |- (1,3)
            node[near end,above]{$P = 1$ if $\Delta H \le 0$}
            node[near end,below]{$P = e^{-\frac{\Delta H}{T_m}}$ if $\Delta H > 0$};
            \draw[-stealth] (-3,-2) |- (1,-3)
            node[xshift=-2.25cm,above]{$P = 0$ if $\Delta H \le 0$}
            node[xshift=-2.25cm,below]{$P = 1 - e^{-\frac{\Delta H}{T_m}}$ if $\Delta H > 0$};
        \end{tikzpicture}
    }
    \caption{Representation of a pixel-copy attempt by one of the pixels of a slow-growing extracellular bacterium to replace one of the surrounding \texttt{Medium} pixels on the two-dimensional square lattice during a MCS. Here, $P$ represents the probability of moving to the cell configuration pointed to by the arrow. The two images on the right-hand side of the figure show two potential outcomes of a pixel-copy attempt, leading to two different potential cell configurations - the top configuration where the pixel-copy attempt is successful; the bottom configuration where it is not.}
    \label{fig:GGH}
\end{figure}
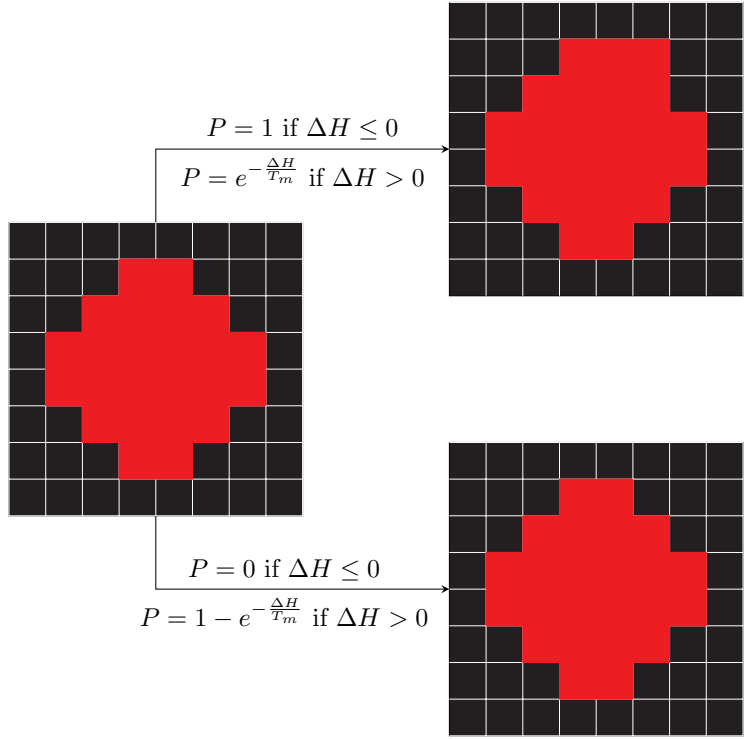

The difference in the Hamiltonian between the current and proposed cell configuration, $\Delta H$, determines whether the proposed cell configuration is accepted or rejected. Specifically, if the generalised cell occupying lattice site $\vec{i}$ is the same as the generalised cell occupying neighbouring lattice site $\vec{j}$, then the cell index at lattice site $\vec{i}$ stays the same. If the generalised cells are different, the probability that the cell index of lattice site $\vec{i}$ changes to the cell index at neighbouring lattice site $\vec{j}$ (or equivalently, the volume of the ``target'' cell that occupies site $\vec{i}$ at the start of the time step decreases by one pixel and the volume of the ``source'' cell that occupies site $\vec{j}$ at the start of the time step increases by one as $\vec{i}$ switches from belonging to the target cell to belonging to the source cell) is as follows:
\begin{equation}
\label{eq:acceptance prob}
    P(\sigma(\vec{i}) \rightarrow \sigma(\vec{j})) = \left \{ \begin{matrix}
    1 & \text{if $\Delta H \le 0$}\\
    f(\Delta H) & \text{if $\Delta H > 0$}.\\
    \end{matrix}
    \right.
\end{equation}
To summarise the implications of Equation (\ref{eq:acceptance prob}), the proposed cell configuration is accepted if the total energy of the system either remains the same or is reduced, as would be expected according to the principle of minimum energy: the higher the energy in a cell configuration, the less probable it is \cite{Glazier2007}. If the total energy of the system would increase with the proposed cell configuration, it is accepted according to an exponentially decreasing function of the difference in the Hamiltonian, $f(\Delta H)$, which is defined as
\begin{equation}
\label{eq:acceptance func}
    f(\Delta H) = e^{-\frac{\Delta H}{T_m}}.
\end{equation}
In Equation (\ref{eq:acceptance func}), $T_m$ is a parameter describing the amplitude of cell membrane fluctuations; it determines the likelihood of accepting energetically unfavourable fluctuations, which helps to avoid getting stuck in energy minima. A representation of an example pixel-copy attempt is shown in Figure \ref{fig:GGH}.\par
Exploring the terms in Equation (\ref{eq:hamiltonian}) individually, the term $E_{\text{adhesion}}$ is a result of the fact that cells are allowed to adhere to each other in our model, and will be more likely to stick to certain cell types than to others. The equation which computes the total energy from these adhesions is
\begin{equation}
\label{eq:adhesion}
    E_{\text{adhesion}} = \sum_{\vec{i} \in \Omega} \sum_{\vec{j} \in \mathcal{N}(\vec{i})} J(\tau(\sigma(\vec{i})),\tau(\sigma(\vec{j})))(1 - \delta(\sigma(\vec{i}),\sigma(\vec{j}))).
\end{equation}
In Equation (\ref{eq:adhesion}), summing over $\vec{j} \in \mathcal{N}(\vec{i})$, the set of neighbouring sites of $\vec{i}$, for all $\vec{i} \in \Omega$, the set of all lattice sites, $J(\tau(\sigma(\vec{i})),\tau(\sigma(\vec{j})))$ is the boundary energy per unit contact area of a pair of cells sticking to one another; the cell at lattice site $\vec{i}$, $\sigma(\vec{i})$ is of type $\tau(\sigma(\vec{i}))$ and the neighbouring cell at lattice site $\vec{j}$, $\sigma(\vec{j})$ is of type $\tau(\sigma(\vec{j}))$. The Kronecker delta function is defined as
\begin{equation}
    \delta(\sigma(\vec{i}),\sigma(\vec{j})) = \left \{
    \begin{matrix}
        1 & \text{if $\sigma(\vec{i}) = \sigma(\vec{j})$,}\\
        0 & \text{otherwise.}\\
    \end{matrix}
    \right .
\end{equation}
This function, $\delta(\sigma(\vec{i}),\sigma(\vec{j}))$, determines whether neighbouring sites are in the same cell or not and consequently we only get a contribution to the adhesion energy if neighbouring sites are parts of different cells.\par
The second term in Equation (\ref{eq:hamiltonian}), $E_{\text{volume}}$, refers to the energy required for cells to maintain their current volume, given a preferred target volume. The equation which computes the total energy arising due to these volume constraints is
\begin{equation}
\label{eq:volume}
    E_{\text{volume}} = \sum_{\sigma} [\lambda^{\text{vol}}_{\sigma}(v(\sigma) - V_t(\sigma))^2].
\end{equation}
In Equation (\ref{eq:volume}), summing over all values of $\sigma$ (that is, summing over all generalised cells), $v(\sigma)$ is the current volume of cell $\sigma$; $V_t(\sigma)$ is the target volume of cell $\sigma$; and $\lambda_{\sigma}^{\text{vol}}$ is the elasticity coefficient, which determines the likelihood of accepting pixel-copy attempts that would move a cell's volume away from its preferred target volume.\par
The third term in Equation (\ref{eq:hamiltonian}), $E_{\text{chem}}$, refers to the energy required for cells of certain cell types to move towards higher concentrations of certain chemical fields via chemotaxis. Specifically, macrophages move chemotactically towards higher concentrations of $C$ (\texttt{Chemokine}). For a macrophage $\Omega_j^{m},j=1,...,N_{m}$, the potential change in energy due to chemotaxis from a pixel-copy attempt, $\Delta E_{\text{chem}}$, is calculated by Equation (\ref{eqn:chemotaxis}):
\begin{equation}
    \Delta E_{\text{chem}} = -\lambda_{\text{chem}}
    [C(\vec{i}) - C(\vec{i'})].
    \label{eqn:chemotaxis}
\end{equation}
In Equation (\ref{eqn:chemotaxis}), $\lambda_{\text{chem}}$ is the Lagrange multiplier determining the strength of the chemotactic attraction (referred to as ``Chemotactic attraction coefficient" in Table \ref{table:Reference parameter set CC3DML}) and $C$ represents the chemokine concentration at the pixel-copy source (destination) pixel $\vec{i'}$ ($\vec{i}$).\par
The final term in Equation (\ref{eq:hamiltonian}), $E_{\text{vec}}$, refers to the energy required for the motility of cells moving in random directions. At each MCS, $(t, t + \Delta t]$, in the model, a random direction, with angle $\theta \in [0,2 \pi)$ to a fixed axis, is chosen for each cell which depends upon its direction at the end of the previous MCS, $t$, such that the direction at the end of the current MCS is calculated according to the following formula:
\begin{equation}
\label{eq:direction}
    \theta(t + \Delta t) \equiv \theta(t) + r \pmod{2\pi}.
\end{equation}
In Equation (\ref{eq:direction}), $r \in [-\pi, \pi]$ is a random number chosen from the von Mises distribution with mode $\mu$, dispersion $\kappa$ and support $[-\pi,\pi]$. If $\mu=0$, the most likely direction by the end of the current MCS is the same as the direction at the end of the previous MCS. The probability density function of angle $r$ given $\mu$ and $\kappa$ is given by
\begin{equation}
\label{eq:von Mises}
    f(r \mid \mu, \kappa) = \frac{\exp(\kappa \cos{(r - \mu)})}{2 \pi I_0(\kappa)},
\end{equation}
where $I_0(\kappa)$ is the modified Bessel function of the first kind of order 0, a scaling constant chosen to ensure the probability distribution integrates to 1. When $\kappa = 0$, this results in a uniform distribution with support $[-\pi,\pi]$; for large $\kappa$, the distribution tends towards a circularly-wrapped normal distribution with mean $\mu$ and variance $1/\kappa$. This was chosen to capture the notion of persistence in cell movement and as a further attempt to reduce lattice anisotropy. To determine the energy that a pixel-copy attempt moving the cell in direction $\theta(t + \Delta t)$ would require, vector quantities in the $x$ and $y$ directions, $\lambda_{\text{vecX}}$ and $\lambda_{\text{vecY}}$, are calculated according to the following trigonometric functions:
\begin{equation}
\label{eq:vecX vecY}
    \lambda_{\text{vecX}} = -F \cos(\theta),\quad\lambda_{\text{vecY}} = -F \sin(\theta).
\end{equation}
In Equation (\ref{eq:vecX vecY}), $F$ is the magnitude of the force required to move the cell. The negative sign reflects the fact that in \texttt{CompuCell3D}, if a cell is more likely to move in the positive direction of an axis then negative energy to move the cell in that direction is more likely to lead to an accepted cell configuration. The change in energy between the current and proposed cell configurations due to random movement, $\Delta E_{\text{vec}}$, is thus calculated as
\begin{equation}
\label{eq:E_vec}
    \Delta E_{\text{vec}} = -(\lambda_{\text{vecX}}\Delta x + \lambda_{\text{vecY}}\Delta y) = -\lambda_{\text{vec}}(\vec{i} - \vec{i'}).
\end{equation}
In Equation (\ref{eq:E_vec}), $\Delta x$ and $\Delta y$ are the displacements of the centres of mass of the cells in the $x$ and $y$ directions; $\vec{i'}$ and $\vec{i}$ are the vector locations of the pixel-copy source and destination pixels; and $\lambda_{\text{vec}}$ is calculated by
\begin{equation}
\label{eq:lambdaVec}
    \lambda_{\text{vec}} = \sqrt{\lambda_{\text{vecX}}^2 + \lambda_{\text{vecY}}^2}.
\end{equation}
%\textbf{Parameter estimation}: each two-dimensional (2-D) pixel in our domain represents $16 \mu m^2$. Each cell (with the exception of \texttt{WALL}) is given the same initial 2-D surface area of $400 \mu m^2$, in-keeping with \cite{Bowness2018}. This means that every cell type we account for in our model is heuristically chosen to be the same size as the largest cell we consider, the macrophage (which has an approximate length of $20 \mu m$ \cite{Krombach1997}). To convert real-time to MCS, we need to take into account that the maximum speed of our generalised cells in \texttt{CompuCell3D} is approximately $0.1 \text{ pixel/MCS}$ \cite{Swat2012}: cells can move slower than this, but not faster. We set the time step according to the fastest-moving cell in our model, the resting macrophage, which moves at a velocity of $1 \mu m/\text{min}$ (this speed was used by \cite{SegoviaJuarez2004}, which based this upon calculations presented in \cite{Webb1996}). Hence one MCS should equal $24 s$. Equivalently, 150 MCS represent 1 hour.
\subsubsection{Oxygen dynamics}
\label{sec:oxygen}
We have included oxygen as a chemical field and two phenotypes of \textit{M. tb} (i.e. two generalised cell types represent \textit{M. tb} depending on its current state). These phenotypes are influenced by the oxygen field and are called ``fast-growing" and ``slow-growing". This has been done to capture the phenotypic changes seen in \textit{M. tb} reverting to a ``dormant" state in the absence of sufficient oxygen, and to study its effect on TB progression.\par
The oxygen chemical field $O:\Omega_T \rightarrow \mathbb{R}$ evolves according to the following reaction-diffusion PDE with zero initial condition and no-flux boundary conditions, which is solved using the forward Euler method for time and a central difference scheme for the spatial discretisation on a square lattice (the same lattice as used for the cell field):
\begin{equation}
\label{eqn:oxygen PDE}
    \frac{\partial O(x,t)}{\partial t} = \nabla \cdot (D_O(x) \nabla O) + \delta (\tau (\sigma (x)),\text{BV}) r_O - \delta (\tau (\sigma (x)),\text{Cell}) \phi_O O.
\end{equation}
In Equation (\ref{eqn:oxygen PDE}), $D_O$ is the oxygen diffusion coefficient, $r_O$ is the uniform supply rate of oxygen through the blood vessels, $\phi_O$ is the consumption rate of oxygen by the bacteria and immune cells, and the Kronecker delta functions are defined as
\begin{equation}
    \delta (\tau (\sigma (x)),\text{BV}) = \left \{
    \begin{matrix}
        1 & \text{if $x \in \Omega^{bv}$,}\\
        0 & \text{otherwise.}\\
    \end{matrix}
    \right .
\end{equation}
\begin{equation}
    \delta (\tau (\sigma (x)),\text{Cell}) = \left \{
    \begin{matrix}
        1 & \text{if $x \in (\Omega^{m} \cup \Omega^b)$,}\\
        0 & \text{otherwise.}\\
    \end{matrix}
    \right .
\end{equation}
When oxygen diffuses through a granuloma, its diffusion rate is reduced \cite{Datta2016,Pienaar2016}. To incorporate this into our model, we reduce the oxygen diffusion constant through cell lattice sites occupied by caseum, infected macrophages and chronically infected macrophages as 
\begin{equation}
\label{eqn:oxygen diffusivity}
    D_O = \left \{
    \begin{matrix}
    \frac{D_O}{D_{OG}} & \text{if $x \in (\Omega^c \cup \Omega^{im} \cup \Omega^{cim})$,}\\
    D_{O} & \text{otherwise.}
    \end{matrix}
    \right.
\end{equation}
\subsubsection{Chemokines}
\label{sec:cytokines}
In our model, we have considered one cytokine, a generic mononuclear chemokine, where \texttt{Chemokine} attracts macrophages to the site of infection.  The macrophages can then perform phagocytosis by engulfing \textit{M. tb}.\par
The chemokine chemical field $C:\Omega_T \rightarrow \mathbb{R}$, evolves according to the following PDE with zero initial condition and no-flux boundary conditions, which is solved using the forward Euler method for time and a central difference scheme for the spatial discretisation on a square lattice (the same lattice as used for the cell field):
\begin{equation}
\label{eqn:chemokine PDE}
    \frac{\partial C(x,t)}{\partial t} = \nabla \cdot (D_{C}  \nabla C) + \delta (\tau (\sigma (x)),\text{IM}\lor\text{CIM}) r_{C} - \eta_{C} C.
\end{equation}
In Equation (\ref{eqn:chemokine PDE}), $D_{C}$ is the chemokine diffusion coefficient, $r_{C}$ is the uniform chemokine secretion rate, $\eta_{C}$ is the chemokine decay rate, IM and CIM represent infected and chronically infected macrophages respectively, $\lor$ is the logical ``or'' operator, and the Kronecker delta function is defined as
\begin{equation}
    \delta (\tau (\sigma (x)),\text{IM}\lor\text{CIM}) = \left \{
    \begin{matrix}
        1 & \text{if $(x,t) \in (\Omega^{im} \cup \Omega^{cim})\times I$,}\\
        0 & \text{otherwise.}\\
    \end{matrix}
    \right .
\end{equation}
%\textbf{Parameter estimation}: the chemokine diffusion coefficient, $D_{Ch}$, is set to $1 \times 10^{-6} \text{ cm}^2/\text{s}$ (obtained from \cite{Francis1997}, which in turn obtained this from \cite{Young1980}) which corresponds to $150 \text{ pixel}^2/\text{MCS}$. Due to instabilities in the forward Euler method, we call the PDE solver 1,000 times per MCS (given by $150/0.15 = 1,000$) to achieve the required diffusion coefficient. The uniform secretion rate, $r_{Ch}$, is set such that the uniform oxygen secretion rate is 2.4 times greater than $r_{Ch}$ (by assumption, in-keeping with \cite{Bowness2018}), which corresponds to $3.28 \times 10^{-3} \text{ mols}/(\text{pixel MCS})$ in \texttt{CompuCell3D} units. The chemokine decay rate, $\eta_{Ch}$, is set equal to $0.347 \text{ h}^{-1}$ (obtained from \cite{Walz1996}), which corresponds to approximately $2.31333... \times 10^{-3}/\text{MCS}$ in \texttt{CompuCell3D}.
\subsubsection{Recruitment of resting macrophages}
Every hour (150 MCS), each blood vessel $\Omega_j^{bv},j=1,...,N_{bv}$ is checked to see if any of the lattice sites in their Moore neigbourhood of order 1 are ``unoccupied", i.e. occupied by the \texttt{Medium} cell type. If so, a new resting macrophage can be recruited on the unoccupied neighbouring lattice site currently being checked. This will happen if
\begin{equation}
\label{eqn:recruitment prob}
    r < 1-\sqrt[nN_{bv}]{1 - M^r_{recr}},
\end{equation}
where $r \sim U(0,1)$ is a uniformly distributed random number between 0 and 1, and $n$ is the total number of times neighbouring lattice sites to blood vessel $\Omega_j^{bv}$ are checked per MCS (as all blood vessels are square $5\times5$ generalised cells, this is equal to 56 - see Figure \ref{fig:recr_sites} for a visual illustration), and $M^r_{recr}$ is the probability of recruiting a resting macrophage per hour. This equation ensures the probability of recruiting a new resting macrophage per hour equals $M^r_{recr}$. The new resting macrophage, which is initially one pixel in volume, is assigned the target volume $V_\sigma$ and elasticity coefficient $\lambda_\sigma^{\text{vol}}$ (which determines how much the cell's volume should be allowed to fluctuate from the target volume). It will subsequently expand its volume to approach $V_\sigma$ over the following MCS.\par
%\textbf{Parameter estimation:} as each blood vessel has size $5 \times 5$, $n$ will equal 56 (as $(4 \times 1) + (8 \times 2) + (12 \times 3) = 56$) (see Figure \ref{fig:recr_sites} for a visual demonstration). $M^r_{recr}$ is obtained from \cite{Bowness2018} (which in turn obtained this from \cite{Cilfone2013}).

\begin{figure}
    \centering
    \begin{tikzpicture} %ADJUST COORDINATES
        \filldraw[fill=White] (-1.75,-1.75) rectangle (1.75,1.75);
        \filldraw[fill=Blue,draw=Black] (-1.25,-1.25) rectangle (1.25,1.25);
        \draw[style=help lines,draw=Black] (-1.25,-1.25) grid[step=0.5cm,xshift=-0.25cm,yshift=-0.25cm] (1.75,1.75);
        \node[draw,rectangle,fill=White,draw=Black] (BL) at (-1.5,-1.5) {1};
        \node[draw,rectangle,fill=White,draw=Black] (2BL) at (-1.5,-1) {2};
        \node[draw,rectangle,fill=White,draw=Black] (3BL) at (-1.5,-0.5) {3};
        \node[draw,rectangle,fill=White,draw=Black] (ML) at (-1.5,0) {3};
        \node[draw,rectangle,fill=White,draw=Black] (3TL) at (-1.5,0.5) {3};
        \node[draw,rectangle,fill=White,draw=Black] (2TL) at (-1.5,1) {2};
        \node[draw,rectangle,fill=White,draw=Black] (TL) at (-1.5,1.5) {1};
        \node[draw,rectangle,fill=White,draw=Black] (T2L) at (-1,1.5) {2};
        \node[draw,rectangle,fill=White,draw=Black] (T3L) at (-0.5,1.5) {3};
        \node[draw,rectangle,fill=White,draw=Black] (TM) at (0,1.5) {3};
        \node[draw,rectangle,fill=White,draw=Black] (T3R) at (0.5,1.5) {3};
        \node[draw,rectangle,fill=White,draw=Black] (T2R) at (1,1.5) {2};
        \node[draw,rectangle,fill=White,draw=Black] (TR) at (1.5,1.5) {1};
        \node[draw,rectangle,fill=White,draw=Black] (2TR) at (1.5,1) {2};
        \node[draw,rectangle,fill=White,draw=Black] (3TR) at (1.5,0.5) {3};
        \node[draw,rectangle,fill=White,draw=Black] (MR) at (1.5,0) {3};
        \node[draw,rectangle,fill=White,draw=Black] (3BR) at (1.5,-0.5) {3};
        \node[draw,rectangle,fill=White,draw=Black] (2BR) at (1.5,-1) {2};
        \node[draw,rectangle,fill=White,draw=Black] (BR) at (1.5,-1.5) {1};
        \node[draw,rectangle,fill=White,draw=Black] (B2R) at (1,-1.5) {2};
        \node[draw,rectangle,fill=White,draw=Black] (B3R) at (0.5,-1.5) {3};
        \node[draw,rectangle,fill=White,draw=Black] (BM) at (0,-1.5) {3};
        \node[draw,rectangle,fill=White,draw=Black] (B3L) at (-0.5,-1.5) {3};
        \node[draw,rectangle,fill=White,draw=Black] (B2L) at (-1,-1.5) {2};
    \end{tikzpicture}
    \caption{Visual illustration that the number of times each recruitment site will be counted for a given blood vessel will sum to 56. The blue $5 \times 5$ square represents a blood vessel in our model.}
    \label{fig:recr_sites}
\end{figure}
\subsubsection{Cell-state transitions}
\label{sec: cell-state transitions}
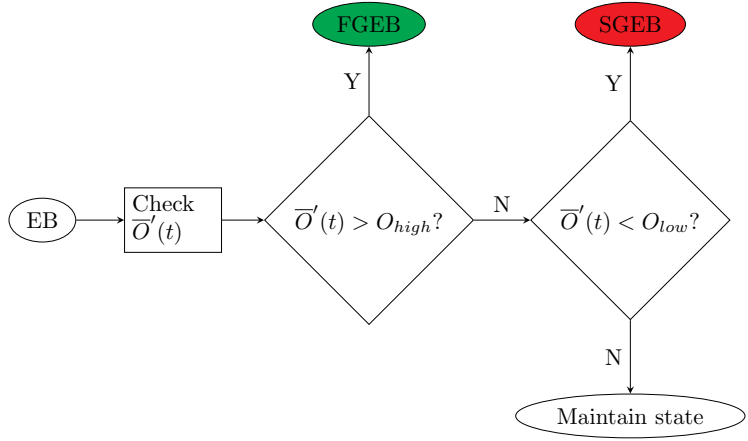
\begin{figure}
    \centering
    \resizebox{10cm}{!}{
        \begin{tikzpicture}
        \node[draw,ellipse,fill=Green] (FGEB) at (0,3){FGEB};
        \node[draw,ellipse,fill=Red] (SGEB) at (4,3){SGEB};
        \node[draw,diamond] (O_high) at (0,0){$\overline{O}'(t) > O_{high}$?};
        \node[draw,diamond] (O_low) at (4,0){$\overline{O}'(t) < O_{low}$?};
        \node[draw,text width = 1.25cm] (process) at (-3,0){Check $\overline{O}'(t)$};
        \node[draw,ellipse] (end) at (4,-3){Maintain state};
        \node[draw,ellipse] (start) at (-5,0){EB};
        \draw[-stealth] (O_high.north) -- (FGEB.south)
        node[midway,left]{Y};
        \draw[-stealth] (O_low.north) -- (SGEB.south)
        node[midway,left]{Y};
        \draw[-stealth] (O_high.east) -- (O_low.west)
        node[midway,above]{N};
        \draw[-stealth] (O_low.south) -- (end.north)
        node[midway,left]{N};
        \draw[-stealth] (process.east) -- (O_high.west);
        \draw[-stealth] (start.east) -- (process.west);
        \end{tikzpicture}
    }
    \caption{Flowchart for extracellular bacteria transition rules. Each extracellular bacterium checks the scaled average oxygen concentration across their volume at time $t$, $\overline{O}'(t)$. If it is above threshold $O_{high}$, it becomes (or remains) fast-growing. Otherwise, if it is below threshold $O_{low}$, it becomes (or remains) slow-growing. Otherwise, it remains in its current state, either fast-growing or slow-growing. Abbreviations: EB, extracellular bacterium; FGEB, fast-growing extracellular bacterium; SGEB, slow-growing extracellular bacterium.}
    \label{fig:FGEB to SGEB}
\end{figure}

\textbf{Rules for extracellular bacteria}: in our model, at each MCS, each \texttt{FGEB} cell $\Omega_j^f, j=1,...,N_f$ and \texttt{SGEB} cell $\Omega_j^s, j=1,...,N_s$ check the average oxygen concentration across their area, scaled between 0 and 100. That is, for every pixel that makes up part of an extracellular bacterium at position $x$ at time $t$, the oxygen concentration $O(x,t)$ is checked; the total oxygen concentration seen across all pixels in a single bacterium is divided by the number of pixels comprising the single bacterium to give the average oxygen concentration across its area at time $t$, $\overline{O}(t)$. The scaled average oxygen concentration across its area, $\overline{O}'(t)$, is then equal to:
\begin{equation}
\label{eq:scaled oxygen}
    \overline{O}'(t) = \frac{\overline{O}(t) - O_{min}}{O_{max} - O_{min}}.
\end{equation}
In Equation (\ref{eq:scaled oxygen}), $O_{min}$ and $O_{max}$ are the minimum and maximum observed oxygen concentrations at any pixel within the domain, respectively. The scaled average oxygen concentration is then compared to two oxygen concentration thresholds, $O_{high}$ and $O_{low}$, with $O_{high} > O_{low}$, to determine if the bacterium changes phenotype. If this scaled average concentration is greater than $O_{high}$, \texttt{SGEB} cells become \texttt{FGEB} cells; if this scaled average concentration is less than $O_{low}$, \texttt{FGEB} cells become \texttt{SGEB} cells; if the concentration is between the two thresholds, the bacterium maintains the same phenotype. This process is illustrated in Figure \ref{fig:FGEB to SGEB}.\par
At the start of the simulation, there is no oxygen present in the domain, so if the bacteria in the initial cluster were to check the oxygen concentration immediately, they would all enter a slow-growing state. To prevent this, we ensure that the bacteria will only check $\overline{O}'(t)$ once they have been alive for 300 MCS, equivalent to 2 hours. This was heuristically chosen to be sufficiently long for a representative oxygen concentration profile to have developed. An alternative implementation could be to introduce the bacteria onto the cell field lattice after the oxygen dynamics have evolved for 300 MCS. However, we have chosen our approach in-keeping with that of the model we are comparing to \cite{Bowness2018}.\par
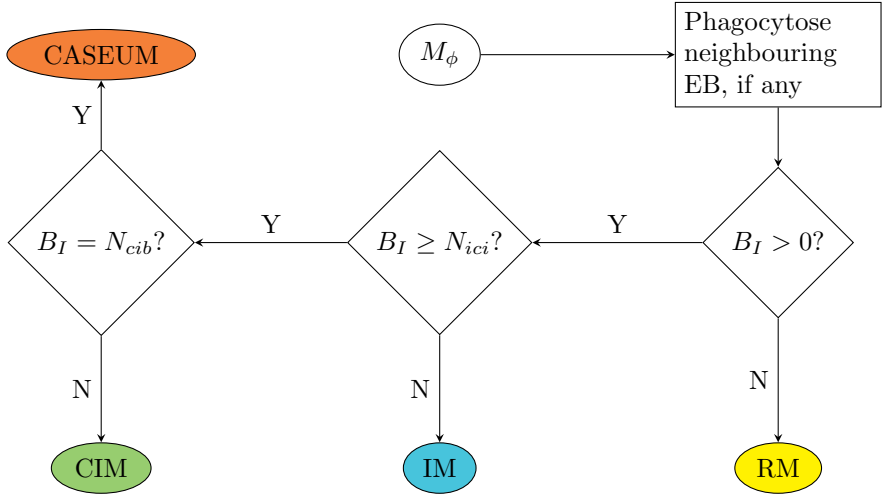
\begin{figure}
    \centering    
        \begin{tikzpicture}
        \node[draw,ellipse] (Mphi) at (-1.5,5.5){$M_{\phi}$};
        \node[draw,text width = 2.5cm] (phagocytosis) at (3,5.5){Phagocytose neighbouring EB, if any};
        \node[draw,diamond] (B_I) at (3,3){$B_I > 0$?};
        \node[draw,diamond] (N_ici) at (-1.5,3){$B_I \ge N_{ici}$?};
        \node[draw,diamond] (N_cib) at (-6,3){$B_I = N_{cib}$?};
        \node[draw,ellipse,fill=Yellow] (RM) at (3,0){RM};
        \node[draw,ellipse,fill=SkyBlue] (IM) at (-1.5,0){IM};
        \node[draw,ellipse,fill=YellowGreen] (CIM) at (-6,0){CIM};
        \node[draw,ellipse,fill=Orange] (CASEUM) at (-6,5.5){CASEUM};
        \draw[-stealth] (Mphi.east) -- (phagocytosis.west);
        \draw[-stealth] (phagocytosis.south) -- (B_I.north);
        \draw[-stealth] (B_I.west) -- (N_ici.east)
        node[midway,above]{Y};
        \draw[-stealth] (B_I.south) -- (RM.north)
        node[midway,left]{N};
        \draw[-stealth] (N_ici.west) -- (N_cib.east)
        node[midway,above]{Y};
        \draw[-stealth] (N_ici.south) -- (IM.north)
        node[midway,left]{N};
        \draw[-stealth] (N_cib.south) -- (CIM.north)
        node[midway,left]{N};
        \draw[-stealth] (N_cib.north) -- (CASEUM.south)
        node[midway,left]{Y};
        \end{tikzpicture}                
    \caption{Flowcharts for macrophage transition rules due to phagocytosis. Here, ``neighbouring'' is defined as being within the von Neumann neighbourhood of range 1 for any of the boundary pixels of the macrophage in question. For more details, see Section \ref{sec: cell-state transitions}. Abbreviations: $M_{\phi}$, macrophage; RM, resting macrophage; IM, infected macrophage; CIM, chronically infected macrophage; EB, extracellular bacteria; $B_I$, intracellular bacterial load.}
    \label{fig:phagocytosis}
\end{figure}

\textbf{Rules for macrophages}: All macrophages are initially resting. At each MCS, every macrophage in our model will check to see if there are extracellular bacteria within the von Neumann neighbourhood of range 1 for each of their boundary pixels, and will phagocytose the bacteria if there are any such bacteria; it will perform this action with certainty. The number of intracellular bacteria now within the macrophage, $B_I$, is calculated. If the macrophage contains no intracellular bacteria, i.e. $B_I = 0$, it remains a resting macrophage. If $0 < B_I < N_{ici}$, it is an infected macrophage; if $N_{ici} \le B_I < N_{cib}$, it is a chronically infected macrophage. If a chronically infected macrophage contains $N_{cib}-1$ bacteria and then phagocytoses another bacterium, it bursts and releases its intracellular bacteria back into the domain, if there is enough space near to the macrophage for the bacteria to move into; if there is not enough space, not all the intracellular bacteria are released, and die with the macrophage. The chronically infected macrophage then becomes caseum after bursting. The flowchart for this process is illustrated in Figure \ref{fig:phagocytosis}.\par
%\textbf{Parameter estimation:} $O_{high}$ and $O_{low}$ were both obtained from \cite{Bowness2018}, where they were chosen to approximately match experimental data. $N_{cib}$ was obtained from \cite{Repasy2013} and \cite{Crowle1986}, where it falls within the ranges of burst sizes. $N_{ici}$ was obtained from \cite{Ray2009}, where it was assumed to be half the size of $N_{cib}$.
\subsubsection{Cell death}
\label{sec:cell death}
All macrophages are initially assigned a lifespan $M_{life}$. Once this lifespan is exceeded, macrophages will die, regardless of current state or intracellular bacterial load. As resting macrophages undergo a controlled form of cell death, they are removed from the model after dying without leaving caseum behind, thereby leaving the cell lattice sites that they previously occupied to be filled by \texttt{Medium}. Infected and chronically infected macrophages will become caseum upon dying naturally (i.e. reaching the end of their assigned lifespan), releasing their intracellular bacterial load.\par
%\textbf{Parameter estimation:} The value for $M_{life}$ was obtained from \cite{VanFurth1973} and \cite{Murphy2008}.
\subsubsection{Bacterial replication}
\label{sec:replication}
Both fast-growing and slow-growing extracellular bacteria will increase in volume over time, provided there is space to expand in size (i.e. there is some part of a bacterium's surface not in contact with any other cells). Given an initial target volume $V_{init}(\sigma)$, each bacterium will increase its target volume every MCS until it reaches a threshold of $2V_{init}(\sigma)$, double its initial target volume (which is set to occur after $Rep_f$ MCS and $Rep_s$ MCS for \texttt{FGEB} cells and \texttt{SGEB} cells, respectively). That is, on a given MCS, each fast-growing extracellular bacterium attempts to increase its target volume by $V_{init}(\sigma)/Rep_f$, and each slow-growing extracellular bacterium attempts to increase its target volume by $V_{init}(\sigma)/Rep_s$, provided there is space to expand in size. Once an extracellular bacterium reaches the threshold of double its initial target volume, it divides into two  extracellular bacteria with each ``child" bacterium having half the volume of the ``parent" bacterium. All other attributes of the ``parent" bacterium are inherited by the ``children" bacteria.\par
%\textbf{Parameter estimation:} The value for $Rep_f$ was obtained from \cite{Shorten2013} and \cite{Beste2009}. The value for $Rep_s$ was obtained from \cite{Beste2009} and \cite{HendonDunn2016}.
\subsection{Parameter estimation}
\label{sec:params}
Each CC3D model has a CC3DML script (which primarily deals with the Glazier-Graner-Hogeweg aspects of the model and the PDE solvers) and a Python script which contains `steppables' (which primarily deals with cell recruitment, cell-state transitions, cell death and bacterial replication; `steppable' is used in CC3D as an alternative name for a class in Python). For the reference parameter values used in the CC3DML script, see Table \ref{table:Reference parameter set CC3DML} (except for the adhesion coefficients, which can be found in Table \ref{table:adhesion}). For the reference parameter values used in the Python steppables, see Table \ref{table:Reference parameter set steppables}. Any parameters which have been estimated using biological data have a justifying reference given alongside them in the table in which they appear. Some parameters could not be estimated in this way, and as such have had to be chosen heuristically. We have chosen these parameter values to ensure consistent behaviour between WHIDM and the CC3D model\footnote{For the parameter values used in WHIDM, see Appendix \ref{appendix:WHIDM params}.}.\par
The adhesion coefficients were chosen to influence the likelihood of cells of one type encountering other cells of either the same type or a different type. In particular, the smallest adhesion coefficient is between \texttt{WALL} cells and \texttt{Medium}, to make the cells of these two types be the most likely to adhere to one another (the smaller the adhesion coefficient, the less energy is required for contact between two cells of the relevant types). As well as this, the joint-largest adhesion coefficients are between \texttt{WALL} cells and any non-\texttt{Medium} cell type other than \texttt{WALL} (that is, between \texttt{WALL} cells and any macrophages, extracellular bacteria, blood vessels or caseum). This was chosen to prevent macrophages sticking to the boundary of the domain, thus enforcing no-flux boundary conditions. The adhesion coefficients between blood vessels and all other cell types were also set to be the joint-largest, to ensure lattice sites that neighbour blood vessels would mostly remain unoccupied and available for macrophage recruitment. The adhesion coefficients between different macrophages were chosen to be large, to prevent them sticking to cells they would not affect. Bacteria were chosen to have small adhesion coefficients with macrophages to encourage phagocytosis. Finally, adhesion coefficients between all non-medium cells and \texttt{Medium} were chosen to be low to encourage cell motility.\par
The volume constraints of cells, chemotactic movement biases of cell movement, and random movement biases of cells can all be strengthened or weakened by altering the scalar values $\lambda_\sigma^{\text{vol}}$, $\lambda_{\text{chem}}$ and $\lambda_{\text{vec}}$ in $E_{\text{volume}}$, $E_{\text{chem}}$ and $E_{\text{vec}}$ (from Equations (\ref{eq:volume}), (\ref{eqn:chemotaxis}) and (\ref{eq:E_vec}), respectively). These are called Lagrange multipliers, and were chosen to make different constraints more or less strict when determining whether a new cell configuration would be accepted or not. Specifically, $\lambda_{\text{chem}}$ was set much larger than $\lambda_{\text{vec}}$ and $\lambda_\sigma^{\text{vol}}$ to give cells that could move chemotactically biased random movement towards higher chemokine concentrations. Similarly, $\lambda_{\text{vec}}$ was set larger than $\lambda_\sigma^{\text{vol}}$ to encourage cells to move randomly in the absence of a chemokine gradient. $\lambda_\sigma^{\text{vol}}$ was increased for infected and chronically infected macrophages on most MCS to be larger than $\lambda_{\text{chem}}$ and $\lambda_{\text{vec}}$, to make them prefer to remain at their current volume, thus `freezing' them in place. $\lambda_\sigma^{\text{vol}}$ is reduced once a sufficient number of MCS have elapsed since the last movement to ensure the correct velocity is maintained (See Table \ref{table:Reference parameter set steppables} for the number of MCS between movements for different types of macrophage). This was done to ensure they move at the correct velocity over the course of the simulation, as velocity could not be explicitly set for cells, so not doing this would allow them to continue moving when they should not be doing so. $\lambda_{\text{vec}}$ could not be set much larger, as an overly large Lagrange multiplier for external potential would cause the cell to fragment trying to move in its randomly chosen direction.\par
Sensible ranges for some of these parameter values have been suggested elsewhere in the literature. For example, setting $\lambda_\sigma^{\text{vol}}$ less than 5 typically causes cells to disappear, whilst setting this parameter greater than 300 `freezes' them in place \cite{Boas2015}. The choice of value for the parameter controlling the average amount of cell membrane fluctuations per MCS, $T_m$, also factors into selecting a reasonable set of values for the Lagrange multipliers. If this value is close to 0, cells will tend to freeze in place regardless of the values for the Lagrange multipliers Whatever value of $T_m$ is chosen, Lagrange multipliers should not be three or more orders of magnitude larger, as this will tend to lead to cells fluctuating out of existence artificially \cite{Scianna2016}. It has been found to be difficult to connect the value of $T_m$ to physical aspects of biological cells \cite{Maree2007}, so we have left $T_m$ at its default value of 10 after consultation with the software developers and factored this into the selection of other parameters relevant to the Glazier-Graner-Hogeweg model.\par
All other parameter values were chosen to reflect estimated instances of containment (defined here as fewer than 10 extracellular bacteria remaining at the end of the simulation) versus dissemination (defined here as at least 10 extracellular bacteria remaining at the end of the simulation), as per \cite{Bowness2018}.
\begin{table} [H]
\centering
\resizebox{12cm}{!}{
\begin{tabular}{|l|l|l|}
\hline
\textbf{Parameter description} & \textbf{Value} & \textbf{Justification}\\
\hline
Lattice dimension ($N \times N \times 1$) & $500 \times 500 \times 1$ & h\\
Cell membrane fluctuation ($T_m$) & 10 & h\\
Voxel copy neighbour order & 5 & \cite{Swat2012}\\
Boundary pixel tracker neighbour order & 1 & h, see Figure \ref{fig:recr_sites}\\
Adhesion neighbour order & 5 & \cite{Swat2012}\\
Simulation duration ($T$) (in MCS) & $3 \times 10^4$ & \cite{Swat2012},\cite{SegoviaJuarez2004},\cite{Webb1996}\\
Chemotactic attraction coefficient ($\lambda_{\text{chem}}$), see (\ref{eqn:chemotaxis}) & $2 \times 10^3$ & h\\
Oxygen diffusion constant ($D_O$), see (\ref{eqn:oxygen PDE}) & $4.62 \times 10^3$ & \cite{Hou2010},\cite{Altman1974}\\
Scalar multiple for oxygen diffusion constant in a granuloma ($\frac{1}{D_{OG}}$) & $\frac{1}{2.7}$ & \cite{Datta2016}\\
Uniform oxygen secretion rate ($r_O$), see (\ref{eqn:oxygen PDE}) & $7.872 \times 10^{-3}$ & \cite{Matzavinos2009}\\
$Ch$ diffusion constant ($D_{C}$), see (\ref{eqn:chemokine PDE}) & 150 & \cite{Francis1997},\cite{Young1980}\\
$C$ decay constant ($\eta_{C}$), see (\ref{eqn:chemokine PDE}) & $2.31333... \times 10^{-3}$ & \cite{Walz1996}\\
Uniform $C$ secretion rate ($r_{C}$), see (\ref{eqn:chemokine PDE}) & $3.28 \times 10^{-3}$ & \cite{Bowness2018}\\
\hline
\end{tabular}
}
\caption{Reference parameter set for CC3DML script, excluding adhesion coefficients. The chemotactic attraction coefficient and directed force coefficient are both set to 0 for any generalised cell that is not a macrophage. Abbreviation: h, heuristically chosen.}
\label{table:Reference parameter set CC3DML}
\end{table}
\begin{table} [H]
%FINISH FILLING IN JUSTIFICATION COLUMN!
    \centering
    \resizebox{14cm}{!}{
    \begin{tabular}{|l|l|l|}
    \hline
         \textbf{Parameter description} & \textbf{Value} & \textbf{Justification}\\
         \hline
         Initial number of fast-growing extracellular bacteria & 6 & \cite{Bowness2018},\cite{CapuanoIII2003}\\
         Initial number of slow-growing extracellular bacteria & 6 & \cite{Bowness2018},\cite{CapuanoIII2003}\\
         Number of blood vessels ($N_{bv}$) & 50 & \cite{Marino2011}\\
         Initial number of \texttt{RM} in the domain ($M^r_{init}$) & 105 & \cite{Wallace1992}\\
         Target volume ($V_\sigma$) (pixels) & 25 & \cite{Bowness2018},\cite{Krombach1997}\\
         Elasticity coefficient ($\lambda^{\text{vol}}_\sigma$) & 10 & h\\
         Directed force coefficient ($\lambda_{\text{vec}}$), see Equation (\ref{eq:lambdaVec}) & 100 & h\\
         Mode of von Mises distribution ($\mu$), see Equation (\ref{eq:von Mises}) & 0 & h\\
         Dispersion of von Mises distribution ($\kappa$), see Equation (\ref{eq:von Mises}) & 99 & h\\
         Oxygen consumption rate of bacteria ($\phi_{O_b}$) & $5.5467 \times 10^{-11}$ & \cite{Sershen2016},\cite{Conkling1982}\\
         Oxygen consumption rate of \texttt{RM} ($\phi_{O_{rm}}$) & $3.067 \times 10^{-13}$ & \cite{Sershen2016},\cite{Conkling1982}\\
         Oxygen consumption rate of \texttt{IM} ($\phi_{O_{im}}$) & $9.2 \times 10^{-13}$ & \cite{Sershen2016},\cite{Conkling1982}\\
         Oxygen consumption rate of \texttt{CIM} ($\phi_{O_{cim}}$) & $1.2267 \times 10^{-12}$ & \cite{Sershen2016},\cite{Conkling1982}\\
         Time between \texttt{FGEB} replications ($Rep_f$) (MCS) & $2.25 - 4.8 \times 10^3$ & \cite{Shorten2013},\cite{Beste2009}\\
         Time between \texttt{SGEB} replications ($Rep_s$) (MCS) & $0.72 - 1.44 \times 10^4$ & \cite{Beste2009},\cite{HendonDunn2016}\\
         Oxygen threshold for $\texttt{FGEB} \rightarrow \texttt{SGEB}$ ($O_{low}$)(\%) & 6 & \cite{Bowness2018}\\
         Oxygen threshold for $\texttt{SGEB} \rightarrow \texttt{FGEB}$ ($O_{high}$)(\%) & 65 & \cite{Bowness2018}\\
         Number of bacteria needed for $\texttt{IM} \rightarrow \texttt{CIM}$ ($N_{ici}$) & 10 & \cite{Ray2009}\\
         Number of bacteria needed for \texttt{CIM} to burst ($N_{cib}$) & 20 & \cite{Repasy2013},\cite{Crowle1986}\\
         Lifespan of \texttt{RM}/\texttt{IM}/\texttt{CIM} ($M_{life}$) (MCS) & 0 - $3.6 \times 10^5$ & \cite{VanFurth1973},\cite{Murphy2008}\\
         Probability of \texttt{RM} recruitment per hour ($M^r_{recr}$) & 0.07 & \cite{Bowness2018},\cite{Cilfone2013}\\
         Extracellular bacterium doubling volume (pixels) & 50 & \cite{Bowness2018},\cite{Krombach1997}\\
         Time between \texttt{RM} movements (MCS) & 1 & \cite{SegoviaJuarez2004},\cite{Webb1996}\\
         Time between \texttt{IM} movements (MCS) & 72 & \cite{SegoviaJuarez2004},\cite{Webb1996}\\
         Time between \texttt{CIM} movements (MCS) & 72 & \cite{SegoviaJuarez2004},\cite{Webb1996}\\
         \hline
    \end{tabular}
    }
    \caption{Reference parameter set for Python steppables script. Where there is a range of values for the parameter (e.g. lifespan), the parameter value is randomly set uniformly in the given range. The unit of measurement for oxygen consumption is $\mu \text{mol}/(\text{pixel} \text{ 10,000 MCS})$. To ensure immune cells move at the correct velocities, the elasticity coefficient is set equal to 100,000 for certain time steps, with the target volume set equal to the cell's current volume to freeze it in place and prevent it from moving when it should not be doing so. Abbreviation: h, heuristically chosen. Cell type indices: \texttt{FGEB}, fast-growing extracellular bacterium; \texttt{SGEB}, slow-growing extracellular bacterium; \texttt{RM}, resting macrophage; \texttt{IM}, infected macrophage; \texttt{CIM}, chronically infected macrophage.}
    \label{table:Reference parameter set steppables}
\end{table}
\begin{table} [H]
    \centering
    \begin{tabular}{|l|rrrrrrrrr|}
    \hline
         & \texttt{Medium} & \texttt{BV} & \texttt{FGEB} & \texttt{SGEB} & \texttt{RM} & \texttt{IM} & \texttt{CIM} & \texttt{CASEUM} & \texttt{WALL}\\
         \hline
         \texttt{Medium} & 10 & 10 & 10 & 10 & 10 & 10 & 10 & 10 & -10\\
         \texttt{BV} &  & 50 & 50 & 50 & 50 & 10 & 10 & 10 & 50\\
         \texttt{FGEB} &  &  & 10 & 10 & 10 & 10 & 10 & 10 & 50\\
         \texttt{SGEB} &  &  &  & 10 & 10 & 10 & 10 & 10 & 50\\
         \texttt{RM} &  &  &  &  & 50 & 50 & 50 & 50 & 50\\
         \texttt{IM} &  &  &  &  &  & 50 & 50 & 50 & 50\\
         \texttt{CIM} &  &  &  &  &  &  & 50 & 50 & 50\\
         \texttt{CASEUM} &  &  &  &  &  &  &  & 10 & 50\\
         \texttt{WALL} &  &  &  &  &  &  &  &  & 10\\
    \hline
    \end{tabular}
    \caption{Triangular array of reference adhesion coefficients. All of these parameter values were heuristically chosen. Cell type indices: \texttt{BV}, blood vessel; \texttt{FGEB}, fast-growing extracellular bacterium; \texttt{SGEB}, slow-growing extracellular bacterium, \texttt{RM}, resting macrophage; \texttt{IM}, infected macrophage; \texttt{CIM}, chronically infected macrophage.}
    \label{table:adhesion}
\end{table}
\subsection{Differences between CC3D model and WHIDM}
\label{sec: comparison to WHIDM}
In this section we outline the differences between the \texttt{CompuCell3D} model and the WHIDM. WHIDM is an agent-based model that can be used to investigate the host-pathogen dynamics of any bacterial infection. As with the CC3D model presented here, bacteria and macrophages are modelled as agents within the framework, and other chemical fields are modelled as the solutions of partial differential equations, solved using the explicit Euler method for time and a central difference scheme for the spatial discretisation. Unlike CC3D, the agents are represented as single points on a lattice, rather than an aggregation of neighbouring lattice sites with a shared index. Updates to cell configurations are not based on energy calculations; rather, events such as cell motility, recruitment, state transitions, phagocytosis and replication are all given certain probabilities of occurring each time step, and will do so over the course of a simulation either after fixed time periods or at a time chosen uniformly at random within a minimum and maximum amount of time, depending upon the input parameters. The WHIDM framework also allowed for T-cells (as agents) and antibiotics (as chemical fields) to be incorporated into simulations; we have chosen not to include these in CC3D due to the short timeframe after initial infection that is being investigated.\par
One of the main differences is the implementation of cell motility. Cells in \texttt{CompuCell3D} require an external force to move, and cannot have their own internally set velocity, whereas cells in WHIDM can move at a certain velocity without any external force being necessary. Furthermore, this velocity is not directly controlled in CC3D: although a cell's velocity tends to be proportional to the external force applied, this force is one of multiple constraints acting upon each cell, along with adhesive forces, attraction to chemoattractants, and volume constraints. The subsequent velocity is therefore affected by the cellular Potts model minimising the energy required for all of these constraints in any potential cell configuration, so it is difficult to get cells to move at the correct velocity. For WHIDM, on the other hand, adhesion and volume constraints are not modelled, and chemotactic bias influences the direction of movement of cells without impacting their speed.\par
The time step of the CC3D model is not the same as that of WHIDM, as the main process that needs to be taken into consideration when determining the model update time step differs between the two approaches. \texttt{CompuCell3D} allows the user to call the PDE solver for each of the chemical fields multiple times per MCS, so the time step can be set according to the fastest-occurring cellular process. WHIDM, on the other hand, runs the PDE solver for each chemical field once per time step, so the time step needs to be set small enough to maintain numerical stability of the solutions to the chemical field PDEs.\par
Additionally, the grid sizes differ: although both models represent a $2 mm \times 2 mm$ patch of lung tissue, WHIDM does this on a $100 \times 100$ grid, compared to the $500 \times 500$ grid for the CC3D model. The primary reason for this is that, if cells are given too small a target volume in \texttt{CompuCell3D}, stochastic fluctuations can lead to unrealistic cell configurations being accepted, in which cells disappear as a result of these fluctuations rather than a real biological process. Theoretically, this is still possible for larger cells, but the probability is substantially reduced so as to make it practically impossible. Consequently, the target volume for cells was set to 25 lattice sites to avoid cells randomly fluctuating out of existence. This value was chosen so that it is larger than the ``temperature" parameter, $T_m$, present in every \texttt{CompuCell3D} model, that dictates the average membrane fluctutaions of a cell for each MCS. This is not incorporated into WHIDM, where cells are single lattice sites in the domain and do not fluctuate in volume or shape. Each grid cell can be occupied by at most one biological element, though, so both models share the volume exclusion property. In WHIDM, lattice spacings, $\delta x$ and $\delta y$, were set equal to the length of the largest cell modelled, the macrophage: each lattice site could contain exactly one macrophage. For the CC3D model, $\delta x$ and $\delta y$ were set so that each $5 \times 5$ square of lattice sites could contain exactly one macrophage with a volume of 25 pixels, so $\delta x$ and $\delta y$ in the CC3D model were set equal to a fifth of the value used for $\delta x$ and $\delta y$ in WHIDM.\par
Replication of \textit{M. tb} was also treated differently by the two modelling approaches. Since \texttt{CompuCell3D} looks to capture biologically realistic behaviour of cells, replication is incorporated as follows: cells increase in volume over time, and divide into two new cells once a threshold target volume is reached. In our CC3D model, each extracellular bacterium would update their target volume every hour, such that they would reach double their original volume after their replication time had elapsed. However, in WHIDM, none of the cells grow in volume over time, always residing on exactly one lattice site, and so a new extracellular bacterium would be created on a free neighbouring lattice site (if one is available) next to another bacterium once the replication time of the other bacterium was reached. If a free neighbouring lattice site is not available, extracellular bacteria in WHIDM change to a ``quiescent'' state and wait for a free neighbouring lattice site to become available before they replicate; we have not included a quiescent extracellular bacteria cell type in our CC3D model.\par
In WHIDM, a granuloma is defined by lattice sites containing caseum in the surrounding neighbourhood. The diffusivity of oxygen is reduced in all lattice sites contained within the granuloma, and any blood vessels contained within the granuloma consequently have their oxygen supply rate reduced. This is approximated in the CC3D model presented here by reducing the diffusion coefficient at any lattice site within a \texttt{CASEUM}, \texttt{IM} or \texttt{CIM} cell, and reducing the oxygen supply at the boundaries of blood vessels that are in direct contact with \texttt{CASEUM}, \texttt{IM} or \texttt{CIM} cells.\par
A comparison of the main features of our CC3D model and the equivalent features of the original WHIDM is shown in Table \ref{table:comparisons}. Many of the features in both of these models are effectively the same, as was intended. However, some of the main differences in the model outputs are caused by certain key parameters in the CC3D model. This is discussed in more detail in Section \ref{section: robustness analysis}.
\begin{table} [H]
\centering
\begin{tabularx}{\linewidth}{
|>{\hsize=0.7\hsize \raggedright\arraybackslash}X
|>{\hsize=1.15\hsize \raggedright\arraybackslash}X
|>{\hsize=1.15\hsize \raggedright\arraybackslash}X|}
\hline
\textbf{Biological process} & \textbf{CC3D} & \textbf{WHIDM}\\
\hline
Cell motility & Motile cells are given  force vectors in the $x$ and $y$ directions and move accordingly & Cells randomly move with uniform probability (in the absence of a chemoattractant) to any lattice site in their Moore neighbourhood of radius 1 at a given rate\\
\hline
Cell adhesion & Cells can adhere to each other & Not modelled\\
\hline
Recruitment of macrophages & A blood vessel can produce a resting macrophage in a free neighbouring lattice site; the macrophage then expands to its target volume in subsequent MCS, provided there is space to do so & A blood vessel can produce a resting macrophage in a free neighbouring lattice site\\
\hline
$\texttt{FGEB} \rightarrow \texttt{SGEB}$ and $\texttt{SGEB} \rightarrow \texttt{FGEB}$ cell-state transitions & The average oxygen concentration across the cell's lattice sites must be below $O_{low}$/above $O_{high}$ & The oxygen concentration at the cell's location must be below $O_{low}$/above $O_{high}$\\
\hline
$\texttt{RM} \rightarrow \texttt{IM}$ cell-state transition & If a resting macrophage encounters extracellular bacteria, it becomes infected & If a resting macrophage encounters extracellular bacteria, it becomes infected\\
\hline
$\texttt{IM} \rightarrow \texttt{CIM}$ cell-state transition & If an infected macrophage cumulatively encounters 10 extracellular bacteria, it becomes chronically infected & If an infected macrophage cumulatively encounters 10 extracellular bacteria, it becomes chronically infected\\
\hline
Chemotaxis & If certain cells sense the chemokine field, they can move towards higher concentrations & If certain cells sense the chemokine field, they can move towards higher concentrations\\
\hline
Cell death & Cells are either deleted or become caseum & Cells are either deleted or become caseum\\
\hline
Bacterial replication & Extracellular bacteria grow to double their size and split in two & A new extracellular bacterium is created next to the original bacterium\\
\hline
Volume constraints & Cells are allowed to fluctuate in volume but have a target volume they prefer to maintain & Cells have a given size and exclude volume\\
\hline
\end{tabularx}
\caption{Comparison of the implementations of the biological processes involved in TB progression between the CC3D model and the within-host infectious disease model.}
\label{table:comparisons}
\end{table}
\section{Results}
\label{section:Results}
Simulations were run for a number of time steps equivalent to 200 hours: this was the period with the most activity in the original model \cite{Bowness2018}. In our CC3D model, this was 30,000 MCS (with each MCS being equivalent to 24s of real time); in WIHMD, this was 200,000 time steps (with each time step equivalent to 3.6s of real time). It should be noted that we are modelling the innate immune response only, and are not considering treatment. Eleven summary statistics were calculated from each simulation, two at the beginning and nine at the end. The two statistics calculated at the beginning of the simulation were the distance from the initial bacterial cluster to the nearest blood vessel, and the number of blood vessels within 0.1 mm of the initial bacterial cluster. The nine statistics calculated at the end of the simulation were the numbers of remaining fast-growing and slow-growing extracellular bacteria; the number of intracellular bacteria; the numbers of remaining resting, infected and chronically infected macrophages; the number of caseous dead cells; the total bacteria in the system (including killed bacteria); and the number of extracellular bacteria killed by the host's immunity (i.e. the number of intracellular bacteria plus the number of killed bacteria). Additionally, the numbers of fast-growing extracellular, slow-growing extracellular and intracellular bacteria were recorded every 150 MCS throughout the simulation.\par
We define containment in our model as the total number of bacteria remaining below a certain threshold at the end of the simulation, as mentioned in Section \ref{sec:params}. In-keeping with \cite{Bowness2018}, we set this threshold equal to 10. If the total number of bacteria is greater than or equal to this threshold at the end of the simulation, we define this as dissemination.\par
The rest of this section is structured as follows: in Section \ref{section:Consistency analysis}, we discuss how many simulations are required to mitigate uncertainty originating from the model's intrinsic stochastic nature so that quantities like mean values and standard deviations can be compared meaningfully; in Section \ref{section: robustness analysis}, we investigate key model parameters to determine how sensitive the CC3D model is to perturbations in their values; in Section \ref{section: results summary}, we summarise the results of our simulations.
\subsection{Consistency analysis}
\label{section:Consistency analysis}
To determine how many simulations should be run in order to be confident that the average results are representative of the model's long-term behaviour, thereby mitigating the model's intrinsic uncertainty resulting from its stochastic nature, we followed the methodology proposed by \citeauthor{Hamis2021} \cite{Hamis2021} and \citeauthor{Alden2013} \cite{Alden2013}. For the CC3D model, we ran $20 \times (1 + 5 + 50 + 100 + 300) = 9,120$ simulations of each model to produce the data samples, one data sample resulting from each simulation. These were then grouped into 20 distributions, each distribution containing $n$ data samples, where $n \in \{1, 5, 50, 100, 300\}$. For WHIDM, we ran $20 \times 300 = 6,000$ simulations instead of 9,120 (see Section \ref{section: WHIDM consistency analysis} for an explanation); beyond this change, we used the same methodology as for CC3D.\par
To compare two distributions, \citeauthor{Vargha2000} \cite{Vargha2000} propose the A-measure of stochastic superiority, which is equal to
\begin{equation}
    A_{12} = P(X_1 > X_2) + 0.5P(X_1 = X_2),
\end{equation}
where $X_1$ and $X_2$ are samples from the two distributions to be compared. The methodology proposed in \cite{Hamis2021,Alden2013} takes a point-estimate of the A-measure, hereafter referred to as the $\hat{A}$-measure, to compare two discrete distributions, $B = \{b_1,...,b_m\}$ and $C = \{c_1,...,c_n\}$, with $m$ and $n$ data samples of random variable $X$, respectively. This is calculated as
\begin{equation}
    \hat{A}_{BC}(X) = \frac{1}{mn} \left ( \sum_{i=1}^m \sum_{j=1}^n \left \{ \mathds{1}_{\{b_i > c_j\}} + \frac{\mathds{1}_{\{b_i = c_j\}}}{2} \right \} \right )
\end{equation}
where $i \in \{1,...,m\}$, $j \in \{1,...,n\}$ and $\mathds{1}$ is the `indicator function', which equals 1 when the event in the subscript curly brackets occurs and 0 otherwise.\par
In each group of 20 distributions, the scaled $\hat{A}$-measures, $\underline{\hat{A}}^n_{1,k'}(X), k' = 2,...,20$ were computed. For data samples $B$ and $C$ of size $n$ and with regard to random variable $X$, the scaled $\hat{A}$-measure is equal to
\begin{equation}
\label{eq:scaled A-hat}
    \underline{\hat{A}}^n_{B,C}(X) = \left \{
    \begin{matrix}
        \hat{A}^n_{B,C}(X) & \text{if $\hat{A}^n_{B,C}(X) \ge 0.5$,}\\
        1 - \hat{A}^n_{B,C}(X) & \text{if $\hat{A}^n_{B,C}(X) < 0.5$.}\\
    \end{matrix}
    \right .
\end{equation}
The $\hat{A}$-measure referred to in (\ref{eq:scaled A-hat}) is equal to
\begin{equation}
    \hat{A}^n_{B,C}(X) = \frac{1}{n^2}\sum_{i = 1}^{n}\sum_{j = 1}^{n}H(b_i - c_j),
\end{equation}
where $H(x)$ is the Heaviside function
\begin{equation}
    H(x) = \left \{
    \begin{matrix}
        1 & \text{if $x > 0$,}\\
        \frac{1}{2} & \text{if $x = 0$,}\\
        0 & \text{if $x < 0$.}
    \end{matrix}
    \right .
\end{equation}
The maximum of these scaled $\hat{A}$-measures, $\max(\underline{\hat{A}}^n_{1,k'}(X))$, was determined for each value of $n$ considered. The minimum amount of runs recommended by the consistency analysis, $n^*$, was then equal to the smallest $n$ such that the maximum scaled $\hat{A}$-measure for that value of $n$ was below some threshold. The threshold was set equal to 0.56, taken from \cite{Hamis2021} as the threshold below which the scaled $\hat{A}$-value indicates a small statistical significance in stochastic differences between samples of a given distribution size.
\subsubsection{CC3D model}
The maximal scaled $\hat{A}$-values for different distribution sizes $n$ from the consistency analysis for the CC3D model are shown in Figure \ref{fig:CC3D maximal scaled A-hat values} (see Figure \ref{fig:CC3D consistency analysis} in Appendix \ref{appendix:CC3D consistency analysis} for all the scaled $\hat{A}$-values). The output variable considered for this analysis was the total amount of \textit{M. tb}, that is, the sum of intracellular and extracellular bacteria, at the end of each simulation. The smallest distribution size for which the statistical significance of the scaled $\hat{A}$-measure is small is $n^* = 300$; this is indicated on Figure \ref{fig:CC3D maximal scaled A-hat values} by the circled value.
\begin{figure}
    \centering
    \subfigure[Maximal scaled $\hat{A}$-values for CC3D for different values of $n$.]{\includegraphics[width=0.45\linewidth]{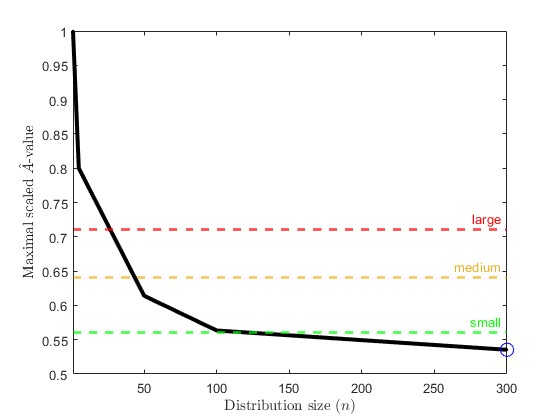}\label{fig:CC3D maximal scaled A-hat values}}
    \hfill
    % REDO FIGURE 6(b) AS JPEG!
    \subfigure[Scaled $\hat{A}$-values for WHIDM ($n=300$).]{\includegraphics[width=0.45\linewidth]{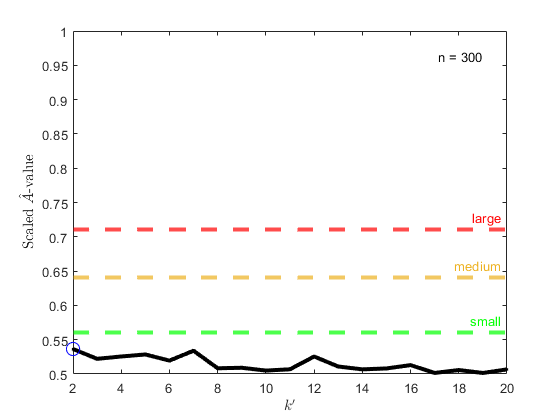}\label{fig:WHIDM n=300}}    
    \caption{Consistency analysis for the CC3D and WHIDM models. The red, yellow and green dashed lines indicate the thresholds below which the scaled $\hat{A}$-values have large, medium and small statistical significance in terms of the stochastic differences between samples of a given distribution size. In Figure \ref{fig:CC3D maximal scaled A-hat values}, we compare five different distribution sizes ($n \in \{1,5,50,100,300\}$); the smallest value of $n$ for which $\max(\hat{A}^n_{1,k'}(X))<0.56$ is circled. In Figure \ref{fig:WHIDM n=300}, we compare 20 distributions of size 300; the maximal scaled $\hat{A}$-value is circled.}
    \label{fig:consistency analysis}
\end{figure}
\subsubsection{WHIDM}
\label{section: WHIDM consistency analysis}
The results of the consistency analysis for the WHIDM are shown in Figure \ref{fig:WHIDM n=300}. As with the consistency analysis of the CC3D model, the output variable considered for this analysis was the total amount of \textit{M. tb}, that is, the sum of intracellular and extracellular bacteria, at the end of each simulation. As we found 300 simulations were sufficient for CC3D, we assumed $n^*$ would equal at least 300 for WHIDM as well and chose this value as the first one to be checked; we would then check larger values if 300 proved insufficient. The smallest distribution size for which the statistical significance of the scaled $\hat{A}$-measure is small is $n^* = 300$, as shown in Figure \ref{fig:WHIDM n=300} by the circled maximal value being below the threshold for small statistical significance in terms of stochastic differences, hence 300 was found to be a sufficient number of simulations to get consistent results from WHIDM.
\subsection{Robustness analysis}
\label{section: robustness analysis}
Having determined what the value of $n^*$ should be, we then conducted a robustness analysis of some of the key parameters of the CC3D model, following the methodology proposed by \citeauthor{Hamis2021} \cite{Hamis2021}. We decided upon the parameters to investigate, and the ranges of values over which they should be perturbed. For each perturbation, we produced a distribution $D_{n^*,p^i_j}$ of $n^*$ data samples of simulations perturbing parameter $p^i$ to value $p^i_j$ and keeping all other parameter values equal to their calibrated values. We then used the $\hat{A}$-measure to compare all distributions $D_{n^*,p^i_j}$ to the distribution $D_{n^*,p^i_C}$ (the distribution corresponding to the calibrated parameter values), in order to determine whether the change in output variable $X$ was statistically significant. Finally, we produced box plots of $X$ against the perturbation values for each parameter to explore the influence of perturbing each parameter.\par
We decided to investigate four parameters: the chemotactic attraction coefficient, bacterium-macrophage contact energy, the dispersion of the von Mises distribution, and the elasticity coefficient (hereafter referred to in this section as $\lambda_{chem}$, $J$, $\kappa$ and $\lambda_{vol}$). These parameters all appear to play an important role in determining the outcome of each simulation, but there is no biological data and limited previous studies in the literature to indicate what values these should take. In particular, $\lambda_{chem}$ will determine how readily immune cells will move to the site of infection and whether they will stay there; $J$ will impact upon the likelihood of phagocytosis occurring; $\kappa$ will dictate over how sharp an angle  macrophages change direction; and $\lambda_{vol}$ will control the elasticity of cells (that is, whether the cells disappear, the cells freeze in place, or something else happens).\par
\begin{figure}
    \centering
    \subfigure[]{\includegraphics[width=0.45\linewidth]{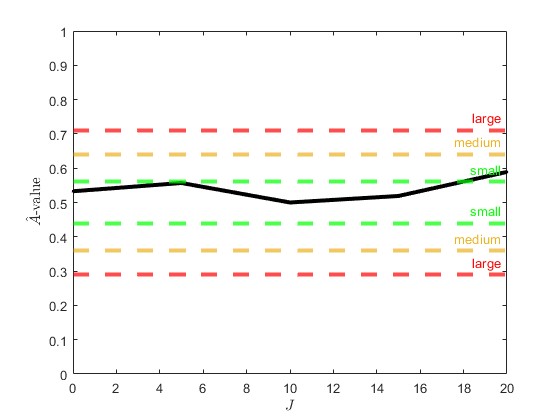}\label{fig:J A-measures}}
    \hfill
    \subfigure[]{\includegraphics[width=0.45\linewidth]{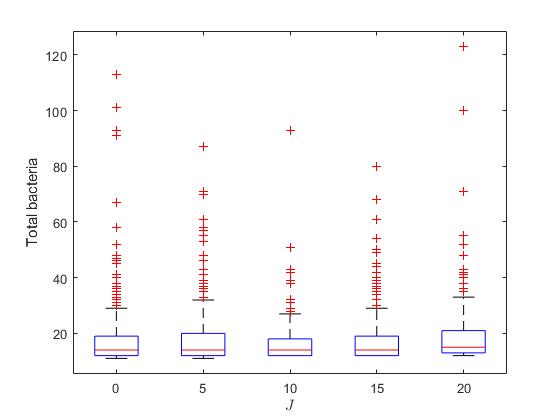}\label{fig:J boxplots}}
    \caption{Summary plots of the robustness analysis regarding  perturbations of the parameter $J$. The values 0, 5, 10, 15 and 20 were considered for this parameter, with $J = 10$ being the calibrated parameter value. The CC3D model was simulated 300 times with $J$ set to the value on the $x$-axis and all other parameter values equal to the ones used to generate the results in Section \ref{section: results summary}. \subref{fig:J A-measures}: $\hat{A}$-measures for perturbations of $J$; \subref{fig:J boxplots}: Box plots of total bacteria for perturbations of $J$.}
    \label{fig:J sensitivity analysis}
\end{figure}
\begin{figure}
    \centering
    \subfigure[]{\includegraphics[width=0.45\linewidth]{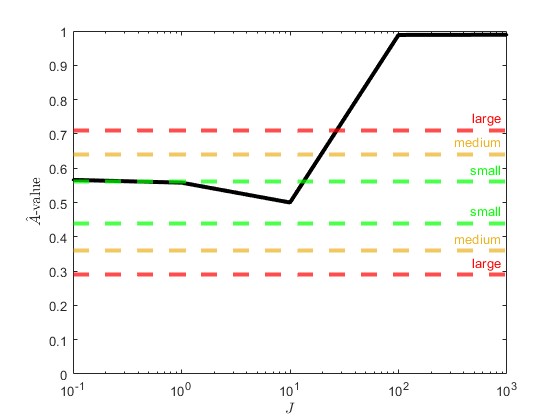}\label{fig:J log A-measures}}
    \hfill
    \subfigure[]{\includegraphics[width=0.45\linewidth]{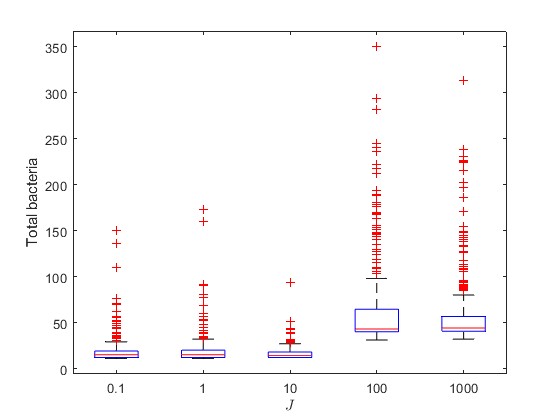}\label{fig:J log boxplots}}
    \caption{Summary plots of the robustness analysis regarding perturbations for a bigger parameter range of the parameter $J$. The values 0.1, 1, 10, 100 and 1000 were considered for this parameter, with $J = 10$ being the calibrated parameter value. The CC3D model was simulated 300 times with $J$ set to the value on the $x$-axis and all other parameter values equal to the ones used to generate the results in Section \ref{section: results summary}. The $x$-axis is on a logarithmic scale. \subref{fig:J log A-measures}: $\hat{A}$-measures for perturbations of $J$; \subref{fig:J log boxplots}: Box plots of total bacteria for perturbations of $J$.}
    \label{fig:J log sensitivity analysis}
\end{figure}
\begin{figure}
    \centering
    \subfigure[]{\includegraphics[width=0.45\linewidth]{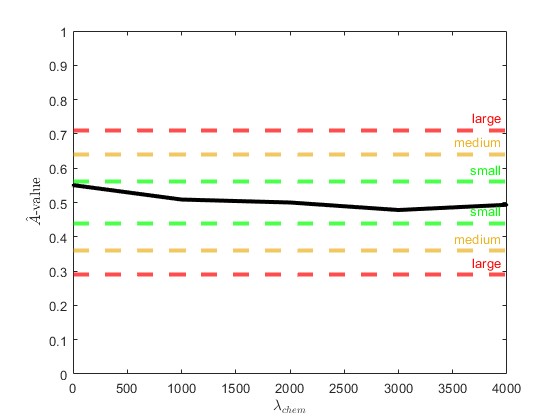}\label{fig:lambda chem A-measures}}
    \hfill
    \subfigure[]{\includegraphics[width=0.45\linewidth]{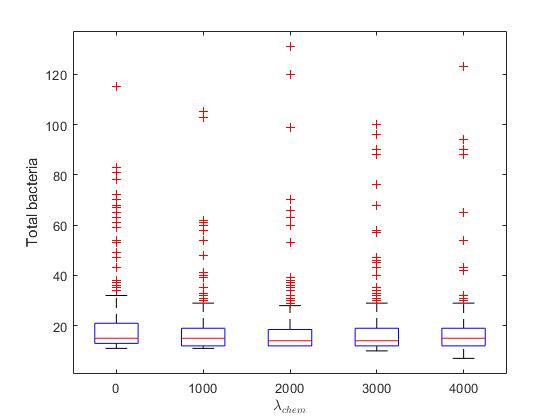}\label{fig:lambda chem boxplots}}
    \caption{Summary plots of the robustness analysis regarding  perturbations of the parameter $\lambda_{chem}$. The values 0, 1000, 2000, 3000 and 4000 were considered for this parameter, with $\lambda_{chem} = 2000$ being the calibrated parameter value. The CC3D model was simulated 300 times with $\lambda_{chem}$ set to the value on the $x$-axis and all other parameter values equal to the ones used to generate the results in Section \ref{section: results summary}. \subref{fig:lambda chem A-measures}: $\hat{A}$-measures for perturbations of $\lambda_{chem}$; \subref{fig:lambda chem boxplots}: Box plots of total bacteria for perturbations of $\lambda_{chem}$.}
    \label{fig:lambda-chem sensitivity analysis}
\end{figure}
\begin{figure}
    \centering
    \subfigure[]{\includegraphics[width=0.45\linewidth]{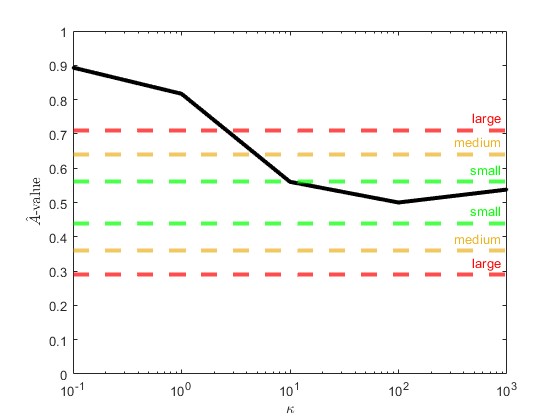}\label{fig:kappa A-measures}}
    \hfill
    \subfigure[]{\includegraphics[width = 0.45\linewidth]{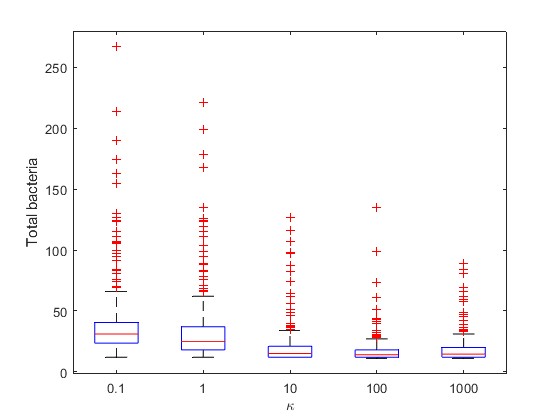}\label{fig:kappa boxplots}}
    \caption{Summary plots of the robustness analysis regarding perturbations of the parameter $\kappa$. The values 0.1, 1, 10, 100 and 1000 were considered for this parameter, with $\kappa = 100$ being the calibrated parameter value (the results were generated with $\kappa = 99 \approx 100$). The CC3D model was simulated 300 times with $\kappa$ set to the value on the $x$-axis and all other parameter values equal to the ones used to generate the results in Section \ref{section: results summary}. The $x$-axis is on a logarithmic scale. \subref{fig:kappa A-measures}: $\hat{A}$-measures for perturbations of $\kappa$; \subref{fig:kappa boxplots}: box plots of total bacteria for perturbations of $\kappa$.}
    \label{fig:kappa sensitivity analysis}
\end{figure}
\begin{figure}
    \centering
    \subfigure[]{\includegraphics[width=0.45\linewidth]{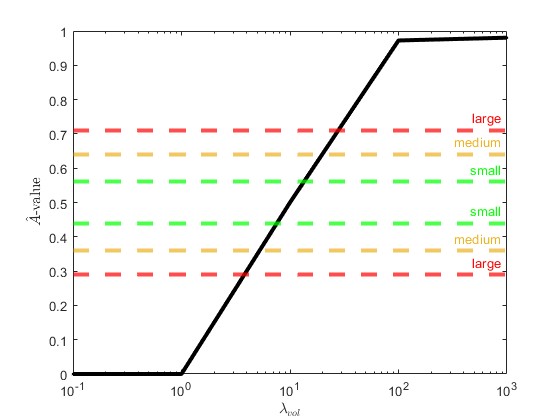}\label{fig:lambda vol A-measures}}
    \hfill
    \subfigure[]{\includegraphics[width=0.45\linewidth]{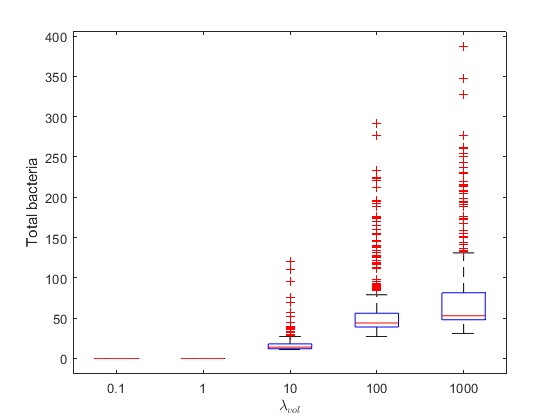}\label{fig:lambda vol boxplots}}
    \caption{Summary plots of the robustness analysis regarding perturbations of the parameter $\lambda_{vol}$. The values 0.1, 1, 10, 100 and 1000 were considered for this parameter, with $\lambda_{vol} = 10$ being the calibrated parameter value. The CC3D model was simulated 300 times with $\lambda_{vol}$ set to the value on the $x$-axis and all other parameter values equal to the ones used to generate the results in Section \ref{section: results summary}. The $x$-axis is on a logarithmic scale. \subref{fig:lambda vol A-measures}: $\hat{A}$-measures for perturbations of $\lambda_{vol}$; \subref{fig:lambda vol boxplots}: box plots of total bacteria for perturbations of $\lambda_{vol}$.}
    \label{fig:lambda vol sensitivity analysis}
\end{figure}

We initially explored five potential values for each parameter. In the case of $J$, we investigated an additional four values after the preliminary results indicated it would be worth exploring the effects of these new values, as the range of values we had initially considered appeared to not be large enough to fully capture the parameter's sensitivity (see Figure \ref{fig:J sensitivity analysis}). The values we considered are listed in Figures \ref{fig:J sensitivity analysis}-\ref{fig:lambda vol sensitivity analysis}. When investigating $\lambda_{chem}$ and $\kappa$, we set the values to be equal for all three cell types affected (i.e. \texttt{RM}, \texttt{IM} and \texttt{CIM}), as we had done when calibrating the model. Similarly, for $J$, we set the parameter values to be equal for any contact energies between an extracellular bacterium (i.e. \texttt{FGEB} and \texttt{SGEB}) and a macrophage (i.e. \texttt{RM}, \texttt{IM} and \texttt{CIM}), as we had done when calibrating the model. Finally, we set $\lambda_{vol}$ to be equal for all cell types affected (\texttt{RM}, \texttt{IM}, \texttt{CIM}, \texttt{FGEB} and \texttt{SGEB}) when determining the model's sensitivity to this parameter.\par
Figure \ref{fig:J sensitivity analysis} shows the results of the robustness analysis for the parameter $J$. As can be seen, the $\hat{A}$-values were not significantly different from one another across most of the perturbations considered, but there is a small statistical difference between the total bacteria numbers for $J = 20$ and the calibrated value of $J = 10$; this suggests the model is sensitive to the value chosen for this parameter beyond a certain threshold. This is to be expected; the greater the value chosen for this parameter, the more energy is required for bacteria and macrophages to adhere to each other, and so the less likely this becomes, and consequently the probability of phagocytosis occurring between a given bacterium and macrophage is reduced. The percentage of simulations that led to dissemination varied between $5\%$ and $8.\dot{6}\%$ for values of $J$ between 0 and 15, but increased to $11.\dot{3}\%$ for $J = 20$. The percentages for the four smaller values considered are broadly in-keeping with the proportions of \textit{M. tb} infections leading to active TB as stated by \citeauthor{Kiazyk2017} \cite{Kiazyk2017}, but the percentage for $J = 20$ is slightly above what would typically be expected, albeit not by much.\par
Subsequently, we decided to try a range of powers of 10 for the values of $J$ to see if increasing the value of this parameter would lead to an increase in the average amount of total bacteria in each simulation and, as a result, an increasing proportion of simulations resulting in dissemination.
As can be seen in Figure \ref{fig:J log sensitivity analysis}, there are large differences in terms of the amount of total bacteria between the simulations generated with the calibrated parameter $J = 10$ and simulations generated with $J$ being set to powers of 10 greater than this. Additionally, the greater the value of $J$, the greater the proportion of simulations resulting in dissemination. In fact, every simulation with $J$ set to either 100 or 1,000 led to dissemination; the number of simulations leading to dissemination varied between 15 and 28 ($5\%$ and $9.\dot{3}\%$) for $J \in \{0.1, 1, 10\}$. This is to be expected: as $J$ increases, the amount of energy needed for contact between a bacterium and a macrophage becomes increasingly large, so the chance of a macrophage being close enough to phagocytose a bacterium decreases.\par
Figure \ref{fig:lambda-chem sensitivity analysis} shows the results of the robustness analysis for the parameter $\lambda_{chem}$. As can be seen, the $\hat{A}$-values were not significantly different from one another across all considered perturbations; this suggests the model is not sensitive to the value chosen for this parameter. The percentage of simulations that led to dissemination decreased monotonically between $10.\dot{6}\%$ ($\lambda_{chem} = 0$) and $6.\dot{6}\%$ ($\lambda_{chem} = 3000$), but slightly increased to $7\%$ for $\lambda_{chem} = 4000$. All of these values are broadly in-keeping with the proportions of infections that lead to active TB stated by \citeauthor{Kiazyk2017} \cite{Kiazyk2017}, and none are statistically significantly different from each other.\par
Although this may seem surprising, this may be due to low concentrations of the chemokine beyond a short distance from the site of infection, in which case an increase in the parameter value will only lead to a smaller increase in guided movement towards the site of infection compared to the smaller parameter value. The chemoattractant diffusion length, $L$ (that is, the length over which the chemoattractant concentration drops to half its value at source) can be calculated as $L=\sqrt{\frac{D_C}{\eta_C}}$ \cite{Bowness2018} \cite{Merks2008}, which equals 255 pixels (rounded to the nearest pixel); if the site of infection is as far away from a macrophage as possible ($\sqrt{500^2+500^2} \approx 707$ pixels away), the chemoattractant concentration should be approximately $2^{707/255} \approx 7$ times smaller than at the site of infection. Given the chemoattractant secretion rate is $3.28 \times 10^{-3}$, the values of $E_{\text{chem}}$ will have order of magnitude of about 1 to 10 when $\lambda_{chem}$ has order of magnitude around 1,000, which is comparatively smaller than other energy values contributing towards the Hamiltonian.\par
Figure \ref{fig:kappa sensitivity analysis} shows the results of the robustness analysis of the parameter $\kappa$. As can be seen, the model is sensitive to values of $\kappa$ lower than 10, where the angle to be added to the current direction of a macrophage is chosen from a less concentrated von Mises distribution (almost a uniform distribution, $U(-\pi,\pi)$, in the case of $\kappa = 0.1$). This may be due to the nature of cell movement in CC3D. A cell effectively moves by copying its index number into lattice sites nearer to the intended target destination and cell types with an alternative index number (for example, \texttt{Medium}) replacing its index number at lattice sites further from the destination. This could cause a more streamlined cell shape to appear if moving more persistently. On the other hand, a sudden change in direction will have the opposite effect, at least temporarily, where the changing of index numbers in lattice sites will cause the cell to become less streamlined and slow it down. This would agree with observations of real biological cells, where faster cells typically move more persistently \cite{Maiuri2015}.\par
As cell speed cannot be directly controlled in the CC3D framework, decreasing $\kappa$ and forcing cells to make sharp changes in direction more often will reduce the average macrophage speed, below the intended realistic speed achieved with the calibrated parameter values, and make them less capable of quickly moving to the site of infection. This is evidenced by the number of simulations leading to dissemination increasing from between 19 and 26 ($6.\dot{3}\%$ and $8.\dot{6}\%$) for $\kappa \in \{10, 100, 1000\}$ to 139 ($46.\dot{3}\%$) for $\kappa = 1$ and increasing even further to 199 ($66.\dot{3}\%$) for $\kappa = 0.1$. This is not to say that the calibrated value of $\kappa$ is the most biologically realistic one - macrophages may move less persistently than we have modelled them as doing. However, the lack of real-world data to suggest an appropriate value for $\kappa$ makes the strong dependence on $\kappa$ a disadvantage of the CC3D model presented here.\par
Figure \ref{fig:lambda vol sensitivity analysis} shows the results of the robustness analysis of the parameter $\lambda_{vol}$. The model is highly sensitive to this parameter: reducing this parameter from its calibrated value by a single order of magnitude leads to no bacteria remaining at the end of the simulation, whereas increasing it by a single order of magnitude causes a large statistically significant increase in the total amount of bacteria observed. For the two smaller values chosen, the cells are unable to maintain their volume and fragment, effectively causing them to `disappear' from the model; even though all the macrophages will disappear, the infection will resolve itself. For the two larger values chosen, the elasticity of the cells is so limited that very few, if any, volume altering pixel copy attempts can be completed, freezing cells in place. In particular, this will mean macrophages do not move towards the site of infection to contain the outbreak, allowing the bacteria to multiply unhindered. This is reflected in the proportion of simulations leading to dissemination: for $\lambda_{vol} \le 1$, no simulations lead to dissemination, whereas for $\lambda_{vol} \ge 100$, essentially all simulations (299 and 300 out of 300 - $99.\dot{6}\%$ and $100\%$ - for $\lambda_{vol} = 100$ and $\lambda_{vol} = 1,000$, respectively) lead to dissemination. Of all the parameters we have investigated, the model appears most sensitive to the choice of this parameter; perturbing the value in either direction from the calibrated value leads to biologically unrealistic scenarios. No real-world data could be found to determine this parameter, intrinsic to the CC3D framework. This is a significant drawback of this approach.\par
All of these parameters impact upon the change in energy between an original configuration of cells and a proposed new one in any given MCS, but they act independently from each other when determining whether the new configuration is to be accepted. However, all of these parameters are impacted by the parameter governing the amplitude of cell membrane fluctuations, $T_{m}$, as discussed in Section \ref{sec:params}. A different value for this parameter would likely impact upon all of the results observed in the robustness analysis: qualitatively, we would expect to see the same patterns, but the actual parameter values needed to generate these results would be scaled according to the choice of $T_{m}$.
\subsection{Summary of simulation results}
\label{section: results summary}
\subsubsection{CC3D model - bacterial growth over time and outcomes}
\begin{figure}
    \centering
    \subfigure[]{\includegraphics[width=0.45\linewidth]{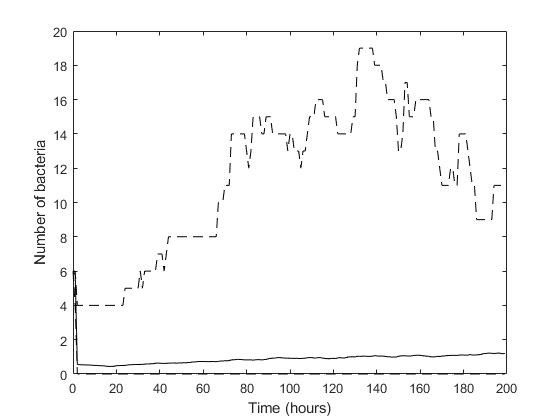}\label{fig:FGEB CC3D}}
    \hfill
    \subfigure[]
    {\includegraphics[width=0.45\linewidth]{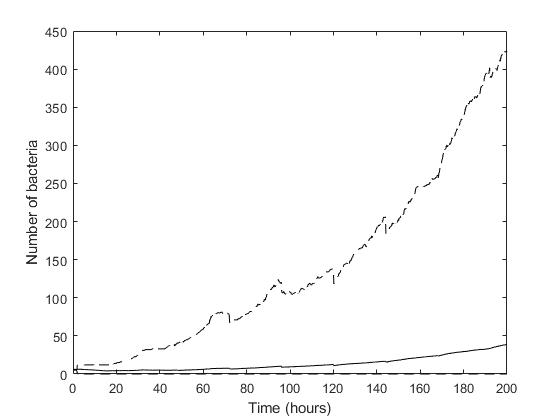}\label{fig:FGEB WHIDM}}
    \subfigure[]{\includegraphics[width=0.45\linewidth]{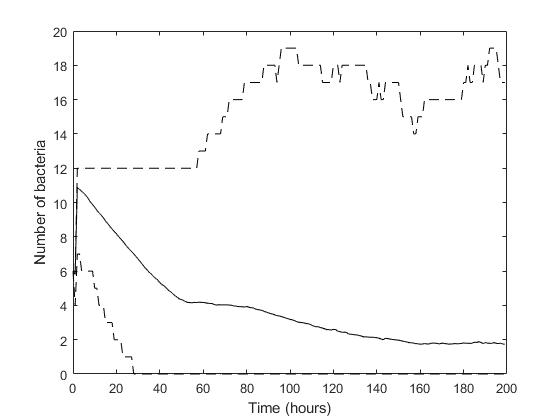}\label{fig:SGEB CC3D}}
    \hfill
    \subfigure[]{\includegraphics[width=0.45\linewidth]{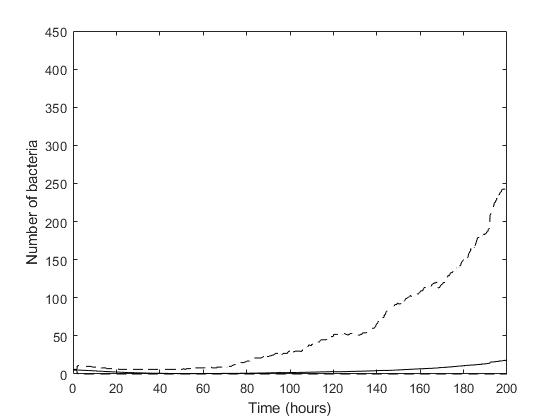}\label{fig:SGEB WHIDM}}
    \subfigure[]
    {\includegraphics[width=0.45\linewidth]{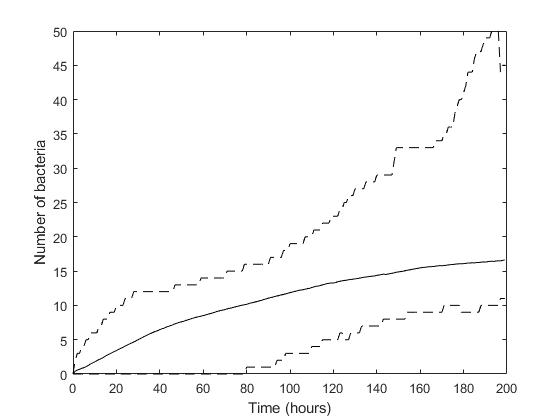}\label{fig:IB CC3D}}
    \hfill
    \subfigure[]{\includegraphics[width=0.5\linewidth]{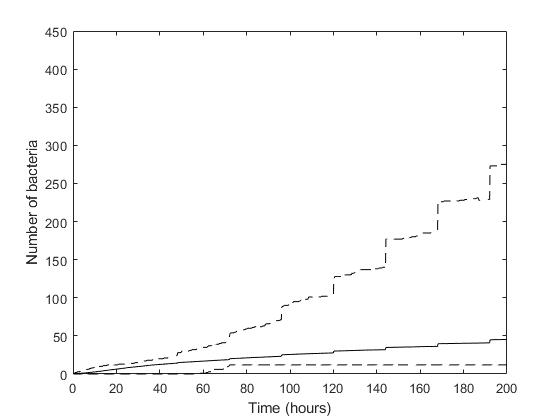}\label{fig:IB WHIDM}}
    \caption{Summary plots of the bacterial populations for all 300 simulations of the CC3D model and WHIDM. The solid lines indicate the mean amounts of \textit{M. tb}. The dashed lines indicate the 95\% confidence intervals. \subref{fig:FGEB CC3D}: fast-growing extracellular bacterial growth over time in CC3D; \subref{fig:FGEB WHIDM}: fast-growing extracellular bacterial growth over time in WHIDM; \subref{fig:SGEB CC3D}: slow-growing extracellular bacterial growth over time in CC3D; \subref{fig:SGEB WHIDM}: slow-growing extracellular bacterial growth over time in WHIDM; \subref{fig:IB CC3D}: intracellular bacterial growth over time in CC3D; \subref{fig:IB WHIDM}: intracellular bacterial growth over time in WHIDM.}
    \label{fig:CC3D WHIDM summary plots}
\end{figure}

Figure \ref{fig:CC3D WHIDM summary plots} summarises the intracellular and extracellular bacteria numbers for 300 simulations of the CC3D model. The macrophages typically phagocytose roughly 15 to 20 \textit{M. tb} on average over the first 200 hours, whilst the extracellular bacteria populations typically decrease to near zero. However, the 95\% confidence intervals show that, on some occasions, the extracellular \textit{M. tb} are able to avoid phagocytosis for the first 200 hours, and slowly start to increase in number over time.\par
\begin{figure}
    \centering
    \subfigure[]{\includegraphics[width=0.45\linewidth]{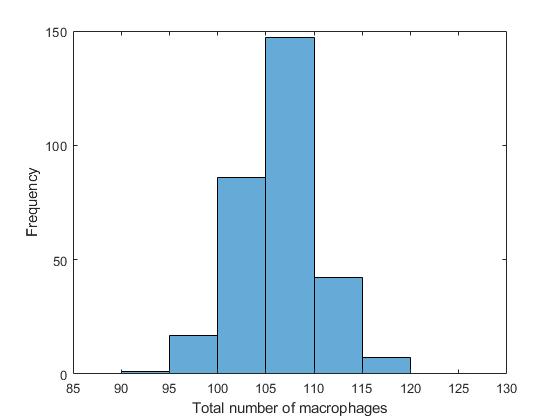}\label{fig:MPhi CC3D}}
    \hfill
    \subfigure[]{\includegraphics[width=0.45\linewidth]{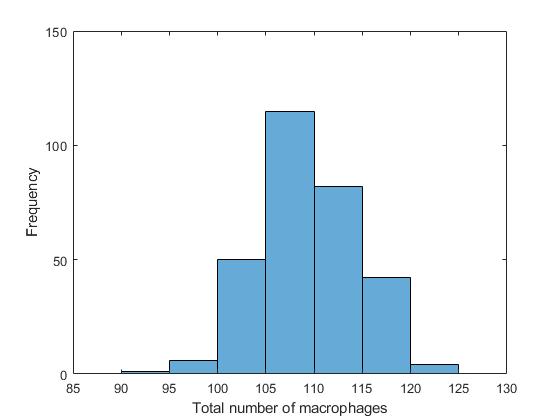}\label{fig:MPhi WHIDM}}
    \caption{Histograms showing the frequencies of the total number of macrophages in the domain at the end of each simulation. \subref{fig:MPhi CC3D}: macrophage numbers in the CC3D model; \subref{fig:MPhi WHIDM}: macrophage numbers in WHIDM.}    
    \label{fig:MPhi}
\end{figure}

Overall, 22 of the 300 simulations ($7.\dot{3}\%$) resulted in dissemination; this is in-keeping with the literature, e.g. \cite{Ahmad2011} \cite{Glaziou2018}. However, we found neither the distance between the randomly-placed initial bacterial cluster and the nearest blood vessel, nor the number of blood vessels within 0.1 mm of the initial bacterial cluster, to be indicative of whether or not containment is achieved. We performed both one-tailed and two-tailed Student's t-tests to see if the differences between the values for containment versus dissemination are statistically significant at that level. They gave p-values greater than 0.05 for both proximity to closest blood vessel and number of blood vessels within a 0.1 mm radius, suggesting the differences are not statistically significant at that level. The median distance between the initial bacterial cluster and the nearest blood vessel was actually shorter for simulations resulting in dissemination (0.035 mm), compared to simulations resulting in containment (0.08 mm). This is surprising, as we would expect macrophages to get to the site of infection and contain the bacteria more easily if they were recruited closer to the site of infection. As can be seen in Figure \ref{fig:MPhi CC3D}, the number of macrophages remains broadly consistent throughout the simulations: there are approximately as many macrophages in the domain at the end of simulations as there are at the start. Therefore, we can discard a drop in macrophage recruitment as a potential reason for these results. Furthermore, the median number of blood vessels within 0.1 mm of the initial bacterial cluster was 1 for both containment and dissemination.\par
One reason for these results is the absence of treatment in the CC3D model: the ability of antibiotics to diffuse effectively to the bacterial cluster helps to explain why proximity to blood vessels has been a key factor in other similar models (e.g. \cite{Bowness2018}, which included treatment), but does not appear to be reflected in the model presented here. Another potentially important aspect is the lack of T-cells or activated macrophages in the CC3D model. We decided to exclude T-cells from our model due to evidence that they don't typically reach the site of infection until after 9-11 days \cite{Urdahl2011}. Consequently, activated macrophages - whose activation is dependent upon the presence of T-cells - were not included either. Finally, it should be noted that, in the CC3D model presented here, we considered the first 200 hours. Long-term simulations that include movement and recruitment of additional immune cells at certain set locations would naturally mean that spatial location would play more of a role in the final outcome. \citeauthor{Bowness2018} also showed the importance of space in modelling within-host TB dynamics in a more general way, in terms of granuloma dynamics and individual heterogeneity, as well as the importance of bacterial phenotypes \cite{Bowness2018}. Some of these effects only become clear after longer periods of simulation, hence why the model we present in this paper has not been able to show these effects as clearly.
\subsubsection{WHIDM - bacterial growth over time and outcomes}
Figure \ref{fig:CC3D WHIDM summary plots} summarises the intracellular and extracellular bacterial numbers over the course of the 300 simulations from the WHIDM. The figures show qualitative agreement between the bacterial growth observed in the WHIDM and that observed in the CC3D model, but some quantitative differences between them: the macrophages typically seem to phagocytose around 45 extracellular bacteria on average over the course of the first 200 hours post-infection, three times as many as in the CC3D model, whilst the extracellular bacteria tend to decrease to a low level. The upper bounds of the 95\% confidence intervals of both extracellular \textit{M. tb} populations reach into the hundreds, showing large outbreaks are occasionally possible with this framework. This is in direct contrast with the results generated by \texttt{CompuCell3D}, where the upper bounds of the 95\% confidence intervals for both extracellular \textit{M. tb} populations never exceeded 20. As can be seen in Figure \ref{fig:MPhi WHIDM}, the total number of macrophages present in the domain at the end of each simulation is similar to those observed in the CC3D model. Overall, 45 out of 300 simulations ($15\%$) resulted in dissemination; this is similar to outcomes observed in reality (which suggest about 5-15\% of \textit{M. tb} infections result in active TB) \cite{Kiazyk2017}. This is more than twice as many disseminations as the results generated by CC3D, but the proportions of disseminations fall within the predicted range of proportions based on outcomes observed in reality for both models. Hence, in the short-term, with antibiotics and adaptive immune response not considered, the outputs of both models show good agreement.
\section{Discussion}
In this section, we summarise the results generated by the two models presented in this work and outline the factors worth bearing in mind when considering whether to use the \texttt{CompuCell3D} framework, and conclude this paper by reviewing our model. This section is structured as follows: Section \ref{section: summary} discusses the implications of this work and the results generated; and Section \ref{section: conclusion} summarises the contents of this paper.
\label{section:Discussion}
\subsection{Summary}
\label{section: summary}
The results generated by the CC3D model are qualitatively similar to those produced by WHIDM, but there are quantitative differences. One of the reasons for this may be due to the nature of CC3D itself. As the modelling framework utilises the cellular Potts approach, the system primarily evolves by minimising the Hamiltonian. As a result, the strong parameter constraints required for the model to run in a physically realistic way, especially the low temperature value, bias the system towards low-energy stable configurations, at the expense of suppressing larger stochastic deviations and outlier behaviours. However, the outcome of \textit{M. tb} infection is volatile: about 1 in 10 infections lead to active disease and potentially very large bacterial populations if granulomas do not form effectively, while the other infections remain latent with small quantities of surviving bacteria if any. Subsequently, this modelling framework is arguably not as suitable for such an infectious disease system, compared to other agent-based modelling alternatives. Having said this, the proportion of disseminations to containments was similar in both modelling frameworks, so it could be argued that both models are useful for exploring the initial host-pathogen interactions after infection from \textit{M. tb}.\par
Qualitative similarities and quantitative differences has previously been observed when \texttt{CompuCell3D} has been used in other areas of mathematical biology. For example, in oncology, a CC3D model of cancer cell growth and invasion was developed and contrasted against an off-lattice alternative model \cite{Andasari2012}. It was found that reducing a key parameter controlling cell-cell adhesion and tumour compactness led to increased tumour invasiveness in both models \cite{Andasari2012}. However, whilst detachment of cells spread radially outwards from a single cell in the lattice-free model, these detachment waves originated from localised groups of cells and spread irregularly in the CC3D model, leading to similar but slightly different results \cite{Andasari2012}. Similarly, although both models presented in this paper are ABMs, the differences in the methodologies give rise to different opportunities and obstacles, depending on the approach chosen.\par
As noted by \citeauthor{Swat2012}, \texttt{CompuCell3D} provides an easy-to-use framework for those less experienced with  software development to create and run models of various biological cellular phenomena \cite{Swat2012}. Previously written code snippets can be added from drop-down menus to both the XML and Python files that cover a wide range of potential model features that the user may wish to add. These may require some editing to match the exact intended usage, but it is usually intuitive how to do so, provided the developer has some prior experience with Python or other similar programming languages.\par
However, further understanding of certain key parameters would be beneficial when using this framework. A lot of the parameters governing the cellular Potts model, in particular the Lagrange multipliers, are abstract quantities. In the future, these parameters could be further refined to more directly match \textit{in vivo} data from the development of real granulomas. Algorithms such as CaliPro can help with finding suitable choices for these parameter values \cite{Joslyn2021}, but this still requires an identification of the range of values to be explored; where to search within the parameter space is unclear without significant trial and error. Additionally, a key advantage of WHIDM over CC3D is the increased level of control of cell movement. In CC3D, each lattice site that comprises a cell and that lattice site's immediate neighbourhood will either remain part of the cell, become part of the cell, or stop being a part of the cell based on the cellular Potts aspect of the model (as opposed to the differential equation solver and ``steppable'' aspects of the model). Although we can set the probabilities to make cells more likely to act a certain way, it is not possible to give cells a definite velocity, for example, as this will vary based upon the stochastic acceptance/rejection of new cell configurations.\par
It is worth noting that, for both the CC3D and WHIDM simulations, we have placed the initial bacterial cluster uniformly at random within the domain, in keeping with the approach presented by \citeauthor{Bowness2018} \cite{Bowness2018}. This may mean that, if this initial cluster is too close to the boundary, a proper granuloma may not be able to form, as immune cells may be unable to get to part of the cluster closest to the boundary. This is a problem that has previously been flagged up by \citeauthor{Bowness2018} in the original study using WHIDM \cite{Bowness2018}. As suggested there, potential solutions may include considering alternatives to no flux boundary conditions, such as periodic boundary conditions, or increasing the size of the domain \cite{Bowness2018}. However, periodic boundary conditions may lead to physically unrealistic granulomas forming, where part of the granuloma is located on the other side of the domain and is not physically connected to the original granuloma. A larger domain size could also prove problematic, as the domain size was chosen to be just large enough to contain a single granuloma. Thus, there is the potential for unrealistically large granulomas to develop in a larger domain. Another alternative, not previously considered by \citeauthor{Bowness2018} in the original study, is to always locate the initial bacterial cluster in the centre of the domain, whilst still randomising the locations of the blood vessels and initial resting macrophages. This should prevent granulomas being unable to form due to the model design, whilst keeping them physically realistic. We did not do so in this study, in order to replicate the methods outlined by \citeauthor{Bowness2018} as closely as possible, but this could be considered in future work.
\subsection{Conclusion}
\label{section: conclusion}
We have presented a within-host model of tuberculosis progression using CC3D. This model captures the host-pathogen interactions within the first 200 hours post-infection from \textit{M. tb} and combines a cellular Potts approach with oxygen and chemokine chemical fields (that evolve according to partial differential equations) and the ability for cells to be recruited, transition between states, replicate and die. This model has produced results that are qualitatively similar to those previously observed in \cite{Bowness2018} - with final average bacteria populations and the containment-to-dissemination ratios not too dissimilar - but quantitatively different in other regards. In that sense, this has proven to be a useful exercise in model cross-validation, as both modelling approaches of TB within-host dynamics are in agreement. However, the upper bound of the 95\% confidence interval for the observed slow-growing extracellular bacterial population is of a different magnitude compared to the same confidence interval generated by the original model \cite{Bowness2018}. Additionally, the distance between the initial cluster and the nearest blood vessel, as well as the number of blood vessels within a 0.1 mm radius of the initial cluster, are not statistically significant in the CC3D model, but had both been significant in previous attempts to model this phenomenon \cite{Bowness2018}; this is likely due to only observing 200 hours, as opposed to 500 days, and subsequently not including either antibiotics treatment or the adaptive immune response. Furthermore, the model has proven to be sensitive to some of the parameters intrinsic to \texttt{CompuCell3D}, without clear evidence for what the values of these parameters should be based on real-world data.\par
We have produced the first TB within-host model using the \texttt{CompuCell3D} framework and shown this framework can model this problem effectively, but researchers will want to consider the issues addressed here when determining whether this framework is right for them to use over other agent-based modelling approaches for similar biological problems. By fine-tuning some of the key parameters, and simulating longer periods of time with the addition of antibiotics and adaptive immunity, we may be able to achieve more comparable results between the CC3D and WHIDM models in future; for now, this is left as future work.
\section*{Acknowledgements}
JD thanks Jakub Mr\'{o}z and Vasileios Alevizos for their participation in the Hackathon team at the 2022 \texttt{CompuCell3D} workshop, which led to the generation of an initial version of the model. JD thanks the instructors of the 2022 \texttt{CompuCell3D} workshop for their support both during and after the workshop. JD thanks Aminat Yetunde Saula for useful discussions regarding the within-host infectious disease model used for comparison against the \texttt{CompuCell3D} model. This work makes use of the Nimbus cloud computer, and the authors gratefully acknowledge the University of Bath's Research Computing Group (\url{doi.org/10.15125/b6cd-s854}) for their support in this work.
\section*{Funding}
This work was supported by the Medical Research Council, United Kingdom [grant number MR/Y010124/1].
\section*{CRediT authorship contribution statement}
\begin{itemize}
    \item \textbf{James W. G. Doran}: Conceptualization, Formal analysis, Investigation, Software, Validation, Visualization, Writing - original draft, Writing - review \& editing.
    \item \textbf{Christopher F. Rowlatt}: Software, Writing - review \& editing.
    \item \textbf{Gibin G. Powathil}: Writing - review \& editing.
    \item \textbf{Ruth Bowness}: Conceptualization, Funding acquisition, Supervision, Writing - review \& editing.
    \item \textbf{Christian A. Yates}: Conceptualization, Supervision, Writing - review \& editing.
\end{itemize}
\section*{Competing interests statement}
The authors declare the following financial interests/personal relationships which may be considered as potential competing interests: Ruth Bowness reports a relationship with Medical Research Council that includes: funding grants (grant number MR/Y010124/1).
\section*{Data accessibility statement}
The CC3D model code can be found at \url{https://github.com/jwgd93/A-model-of-tuberculosis-progression-using-CompuCell3D}; the WHIDM code can be found at \url{https://github.com/Ruth-Bowness-Group/A-model-of-tuberculosis-progression-using-CompuCell3D-WHIDM}. The data presented in this article can be found at the following links: \url{https://doi.org/10.5281/zenodo.18672018} for the CC3D consistency analysis data; \url{https://doi.org/10.5281/zenodo.18672594} for the WHIDM consistency analysis data; \url{https://doi.org/10.5281/zenodo.18672733} for the CC3D robustness analysis data; \url{https://doi.org/10.5281/zenodo.18673176} for the CC3D simulation results; \url{https://doi.org/10.5281/zenodo.18664196} for the WHIDM simulation results. These links will become live after publication.
\begin{appendices}
\section{Consistency analysis}
\begin{figure}[H]
    \centering
    \subfigure[$n = 1$]{\includegraphics[width=0.3\linewidth]{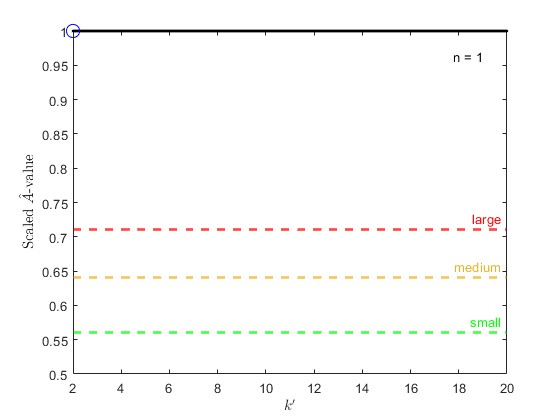}}
    \hfill
    \subfigure[$n = 5$]{\includegraphics[width=0.3\linewidth]{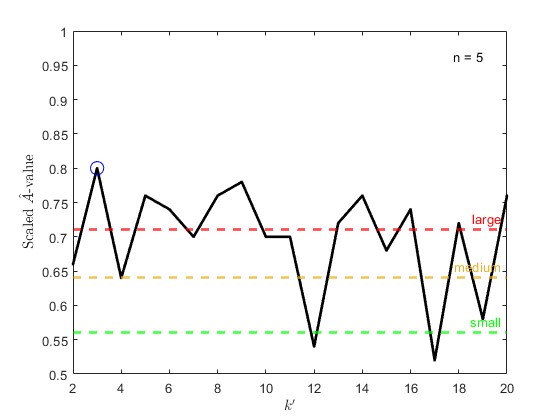}}
    \hfill
    \subfigure[$n = 50$]{\includegraphics[width=0.3\linewidth]{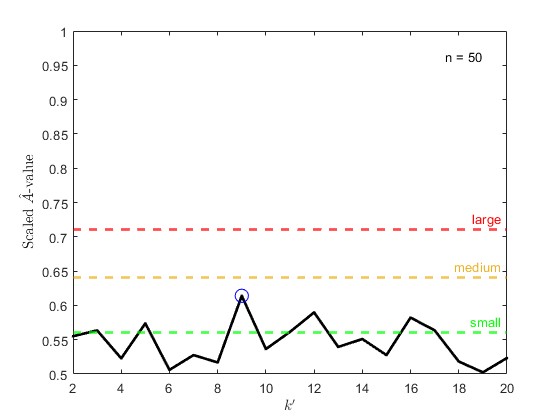}}
    \subfigure[$n = 100$]{\includegraphics[width=0.3\linewidth]{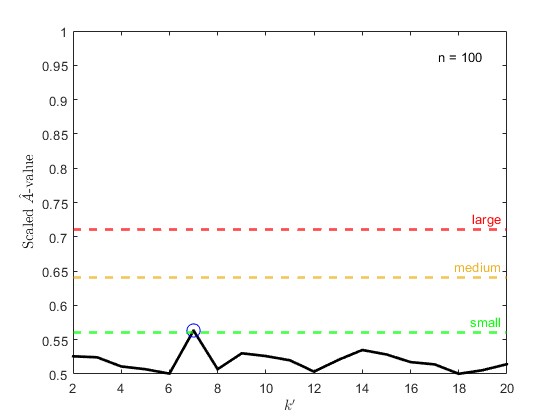}}
    \subfigure[$n = 300$]{\includegraphics[width=0.3\linewidth]{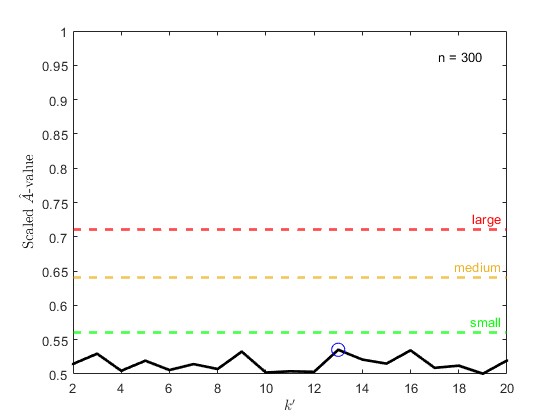}}
    \caption{Consistency analysis for the CC3D model. The scaled $\hat{A}$-values for different distribution sizes are shown. The red, yellow and green dashed lines indicate the thresholds below which the scaled $\hat{A}$-values have large, medium and small statistical significance in terms of the stochastic differences between samples of a given distribution size. The maximal value is circled.}
    \label{fig:CC3D consistency analysis}
\end{figure}

This appendix shows the outcomes of the consistency analysis for the CC3D model. As can be seen in Figure \ref{fig:CC3D consistency analysis}, the smallest distribution size for which the statistical significance of the scaled $\hat{A}$-measure is small is $n^*=300$ (that is, $\max(\hat{A}^{n^*}_{1,k'}(X))<0.56$ and $\max(\hat{A}^n_{1,k'}(X)) \ge 0.56 \forall n < n^*$).
\label{appendix:CC3D consistency analysis}
\newpage
\section{WHIDM parameters}
\label{appendix:WHIDM params}
The following table lists all the parameter values used in the WHIDM simulations.
\begin{table} [H]
\centering
\resizebox{12cm}{!}{
\begin{tabular}{|l|l|l|}
\hline
\textbf{Parameter description} & \textbf{Value}\\
\hline
Output interval (h) & 0.2\\
Simulation duration (h) & 200\\
Timestep (h) & 0.001\\
Grid shape & $100 \times 100$\\
Grid spacing (cm) & 0.002\\
Maximum neighbourhood size & 5\\
Granuloma neighbourhood size & 2\\
Oxygen diffusion coefficient ($\text{cm}^2\text{ h}^{-1}$) & 0.11088\\
Oxygen decay ($\text{h}^{-1}$) & 0\\
Oxygen granuloma diffusion reduction & 2.7\\
Chemokine diffusion coefficient ($\text{cm}^2\text{ h}^{-1}$) & 0.0036\\
Chemokine decay ($\text{h}^{-1}$) & 0.347\\
Number of blood vessels & 50\\
Oxygen secretion rate ($\text{mol h}^{-1}$) & 29.52\\
Oxygen granuloma secretion rate reduction & 4\\
Resting macrophage recruitment rate per blood vessel & $1.45 \times 10^{-6}$\\
Macrophage area ($\text{cm}^2$) & $3.14 \times 10^{-6}$\\
Macrophage lifespan (h) & $1200 \pm 1200$\\
Macrophage maximum neighbourhood size & 3\\
Initial number of resting macrophages & 105\\
Resting macrophage oxygen uptake ($\text{h}^{-1}$) & $1.15 \times 10^{-13}$\\
Infected macrophage oxygen uptake ($\text{h}^{-1}$) & $3.45 \times 10^{-13}$\\
Infected macrophage chemokine secretion rate ($\text{nM h}^{-1}$) & 0.0126\\
Chronically infected macrophage oxygen uptake ($\text{h}^{-1}$) & $4.6 \times 10^{-13}$\\
Chronically infected macrophage chemokine secretion rate ($\text{nM h}^{-1}$) & 0.0126\\
Time interval for resting macrophage movement (h) & 0.333\\
Time interval for infected macrophage movement (h) & 24\\
Time interval for chronically infected macrophage movement (h) & 24\\
Macrophage chemotactic migration bias & 19.4\\
Macrophage chemotactic migration weight & 1\\
Phagocytosis probability & 1\\
Number of bacteria needed for $\texttt{RM} \rightarrow \texttt{IM}$ & 1\\
Probability of $\texttt{RM} \rightarrow \texttt{IM}$ after threshold reached & 1\\
Number of bacteria needed for $\texttt{IM} \rightarrow \texttt{CIM}$ & 10\\
Probability of $\texttt{IM} \rightarrow \texttt{CIM}$ after threshold reached & 1\\
Number of bacteria needed for $\texttt{CIM}$ to burst & 20\\
Probability of $\texttt{CIM}$ bursting after threshold reached & 1\\
Maximum neighbourhood size for distributing bacteria after $\texttt{CIM}$ burst & 3\\
Bacterium are ($\text{cm}^2$) & $3.14 \times 10^{-6}$\\
Bacterium lifespan (h) & 200\\
Initial number of fast-growing extracellular bacteria & 6\\
Initial number of slow-growing extracellular bacteria & 6\\
Fast-growing extracellular bacterium oxygen uptake rate ($\text{h}^{-1}$) & $2.08 \times 10^{-11}$\\
Slow-growing extracellular bacterium oxygen uptake rate ($\text{h}^{-1}$) & $2.08 \times 10^{-11}$\\
Extracellular bacteria state switch delay (h) & 2\\
Oxygen threshold for $\texttt{FGEB} \rightarrow \texttt{SGEB}$ (\%) & 6\\
Oxygen threshold for $\texttt{SGEB} \rightarrow \texttt{FGEB}$ (\%) & 65\\
Fast-growing extracellular bacterium replication rate (h) & 15-32\\
Slow-growing extracellular bacterium replication rate (h) & 48-96\\
Extracellular bacterium replication maximum neighbourhood size & 3\\
\hline
\end{tabular}
}
\caption{Reference parameter set for WHIDM. All simulations are conducted in 2D, hence no third dimension length is specified. The initial extracellular bacteria are deposited in a square region with a width and height of 6 grid spaces, with the location chosen uniformly at random. Abbreviations: $\texttt{RM}$, resting macrophage; $\texttt{IM}$, infected macrophage; $\texttt{CIM}$, chronically infected macrophage; $\texttt{FGEB}$, fast-growing extracellular bacterium; $\texttt{SGEB}$, slow-growing extracellular bacterium.}
\label{table:Reference parameter set WHIDM}
\end{table}
\end{appendices}
\newpage
\raggedright \printbibliography
\end{document}